\RequirePackage[mathlines,running]{lineno}
\documentclass[twocolumn,trackchanges]{aastex631}

\usepackage[normalem]{ulem}

\usepackage{amssymb,amsmath}
\usepackage{multirow,comment,enumerate,bm}
\usepackage[nolist,nohyperlinks]{acronym}

\received{2024 March 12}
\revised{2024 May 24}
\accepted{2024 May 29}
\published{2024 August 13}

\newcommand{\secref}[1]{\S\ref{#1}}

\newcommand{\posydon}{\ensuremath{\texttt{POSYDON}}}
\newcommand{\mesa}{\ensuremath{\texttt{MESA}}}

\newcommand{\CIERA}{Center for Interdisciplinary Exploration and Research in Astrophysics (CIERA), 1800 Sherman, Evanston, IL 60201, USA}
\newcommand{\NUPhysAstro}{Department of Physics \& Astronomy, Northwestern University, 2145 Sheridan Road, Evanston, IL 60208, USA}

\newcommand{\RV}{\ifmmode {{\rm RV}}\else RV \fi}

\newcommand{\rot}{ \ensuremath{{\omega/\omega_\mathrm{crit}}} }
\begin{acronym}[MPC]
\acro{BBH}{binary black hole}
\acro{BH}{black hole}
\acro{NS}{neutron star}
\acroplural{NS}[NSs]{neutron stars}

\acro{CO}{compact object}

\acro{ZAMS}{Zero Age Main Sequence}
\acro{TAMS}{Terminal Age Main Sequence}

\acro{SFH}{star formation history}

\acro{CE}{common envelope}
\acro{SN}{supernova}
\acroplural{SN}[SNe]{supernovae}
\acro{RLO}{Roche-lobe overflow}
\acro{MT}{mass transfer}

\acro{HMXB}{high-mass X-ray binary}
\acroplural{HMXB}[HMXBs]{high-mass X-ray binaries}
\acro{BeXB}[BeXB]{Be X-ray binary}
\acroplural{BeXB}[BeXBs]{Be X-ray binaries}

\acro{AL}{active learning}
\acro{ML}{machine learning}
\acro{RBF}{radial basis function}
\acro{PTMCMC}{parallel tempered Markov-chain Monte Carlo}

\acro{mesa}[\ensuremath{\texttt{MESA}}]{Modules for Experiments in Stellar Astrophysics}
\acro{posydon}[\ensuremath{\texttt{POSYDON}}]{POpulation SYnthesis  with Detailed binary-evolution simulatiONs}
\acro{bse}[\ensuremath{\texttt{BSE}}]{Binary Stellar Evolution}
\acro{sse}[SSE]{single star evolution}

\acro{3D}{three-dimensional}
\acro{1D}{one-dimensional}
\end{acronym}

\begin{document}

\title{To Be or not to Be: the role of rotation in modeling Galactic Be X-ray Binaries}

\author[0000-0003-4474-6528]{Kyle~Akira~Rocha}
\affiliation{\NUPhysAstro{}}
\affiliation{\CIERA{}}
\email{kylerocha2024@u.northwestern.edu}

\author[0000-0001-9236-5469]{Vicky Kalogera}
\affiliation{\NUPhysAstro{}}
\affiliation{\CIERA{}}

\author[0000-0002-2077-4914]{Zoheyr Doctor}
\affiliation{\CIERA{}}
\affiliation{\NUPhysAstro{}}

\author[0000-0001-5261-3923]{Jeff~J.~Andrews}
\affiliation{Department of Physics, University of Florida, 2001 Museum Rd, Gainesville, FL 32611, USA}

\author[0000-0001-9037-6180]{Meng\,Sun}
\affiliation{\CIERA{}}

\author[0000-0001-6692-6410]{Seth\,Gossage}
\affiliation{\CIERA{}}

\author[0000-0002-3439-0321]{Simone\,S.\,Bavera}
\affiliation{Département d’Astronomie, Université de Genève, Chemin Pegasi 51, CH-1290 Versoix, Switzerland}
\affiliation{Gravitational Wave Science Center (GWSC), Université de Genève, CH1211 Geneva, Switzerland}

\author[0000-0003-1474-1523]{Tassos\,Fragos}
\affiliation{Département d’Astronomie, Université de Genève, Chemin Pegasi 51, CH-1290 Versoix, Switzerland}
\affiliation{Gravitational Wave Science Center (GWSC), Université de Genève, CH1211 Geneva, Switzerland}

\author[0000-0003-3684-964X]{Konstantinos\,Kovlakas}
\affiliation{Institute of Space Sciences (ICE, CSIC), Campus UAB, Carrer de Magrans, 08193 Barcelona, Spain}
\affiliation{Institut d’Estudis Espacials de Catalunya (IEEC), Carrer Gran Capit\`a, 08034 Barcelona, Spain}

\author[0000-0001-9331-0400]{Matthias\,U.\,Kruckow}
\affiliation{Département d’Astronomie, Université de Genève, Chemin Pegasi 51, CH-1290 Versoix, Switzerland}
\affiliation{Gravitational Wave Science Center (GWSC), Université de Genève, CH1211 Geneva, Switzerland}

\author[0000-0003-4260-960X]{Devina\,Misra}
\affiliation{Département d’Astronomie, Université de Genève, Chemin Pegasi 51, CH-1290 Versoix, Switzerland}
\affiliation{Institutt for Fysikk, Norwegian University of Science and Technology, Trondheim, Norway}

\author[0000-0003-1749-6295]{Philipp\,M.\,Srivastava}
\affiliation{\CIERA{}}
\affiliation{Electrical and Computer Engineering, Northwestern University, 2145 Sheridan Road, Evanston, IL 60208, USA}

\author[0000-0002-0031-3029]{Zepei\,Xing}
\affiliation{Département d’Astronomie, Université de Genève, Chemin Pegasi 51, CH-1290 Versoix, Switzerland}
\affiliation{Gravitational Wave Science Center (GWSC), Université de Genève, CH1211 Geneva, Switzerland}

\author[0000-0002-7464-498X]{Emmanouil\,Zapartas}
\affiliation{Institute of Astrophysics, FORTH, N. Plastira 100,  Heraklion, 70013, Greece}

\begin{abstract}
Be X-ray binaries (Be-XRBs) are one of the largest subclasses of high-mass X-ray binaries, comprised
of a rapidly rotating Be star and neutron star companion in an eccentric orbit, intermittently
accreting material from a decretion disk around the donor.
Originating from binary stellar evolution, Be-XRBs are of significant interest to binary population synthesis (BPS) studies, encapsulating the physics of supernovae, common envelope, and \ac{MT}. 
Using the state-of-the-art BPS code, \posydon{}, which relies on pre-computed grids of detailed binary stellar evolution models, we investigate the Galactic Be-XRB population. 
\posydon{} incorporates stellar rotation self-consistently during \ac{MT} phases, enabling detailed examination of the rotational distribution of Be stars in multiple phases of evolution. 
Our fiducial BPS and Be-XRB model align well with the orbital properties of Galactic Be-XRBs, emphasizing the role of rotational constraints. 
Our modeling reveals a rapidly rotating population ($\rot{} \gtrsim 0.3$) of Be-XRB-like systems with a strong peak at intermediate rotation rates ($\rot{} \simeq 0.6$) in close alignment with observations.
All Be-XRBs undergo a \ac{MT} phase before the first compact object forms, with over half experiencing a second \ac{MT} phase from a stripped helium companion (Case BB). 
Computing rotationally-limited \ac{MT} efficiencies and applying them to our population, we derive a physically motivated MT efficiency distribution, finding that most Be-XRBs have undergone highly non-conservative \ac{MT} ($\bar{\beta}_\mathrm{rot} \simeq 0.05$). 
Our study underscores the importance of detailed angular momentum modeling during \ac{MT} in interpreting Be-XRB populations, emphasizing this population as a key probe for the stability and efficiency of \ac{MT} in interacting binaries.
\end{abstract}


\section{Introduction}
\label{sec:intro}
Classical Be stars are understood to be rapidly rotating spectral-type B stars that have been observed with Balmer emission lines 
and excess infrared emission,  attributed to a circumstellar, rotationally supported \textit{decretion} disk emanating from the star \citep{Rivinius2013AARv}.
The idea that rapid rotation of Be stars is linked to the formation of the decretion disk, first proposed by \cite{Struve1931ApJ}, remains a leading theory for the so-called Be-phenomenon, as observations consistently show the Be population as a whole is highly spinning \citep[e.g.,][]{Yudin2001AA,Rivinius2013AARv,Zorec2016AA,Balona-Ozuyar2021ApJ,Kamann2023MNRAS}. 
Although the true rotational distribution of Be stars is in and of itself an active area of inquiry \citep[e.g.][]{Townsend2004MNRAS,Yudin2001AA,Rivinius2013AARv,Balona-Ozuyar2021ApJ,Abdul-Masih2023AA}.
While there have been extensive studies into the evolution of the disk in isolated Be stars and under the influence of a companion in a binary, the exact mechanism(s) responsible for launching the disk remain a topic of debate \citep{Okazaki1997AA,Okazaki2011PASJ,Martin2011MNRAS,Martin2014ApJ,Panoglou2016MNRAS,Cyr2017MNRAS,Brown2018MNRAS,Panoglou2018MNRAS,Brown2019MNRAS,Cyr2020MNRAS,Franchini2019ApJ,Franchini2021ApJ,Suffak2022MNRAS}.
Theories for decretion disk formation include non-radial stellar pulsations \citep[][]{Ressler2021MNRAS} and/or small scale surface magnetic fields \citep[][]{NixonPringle2020}, but all require rapid rotation to effectively reduce the surface gravity at the stellar equator for material to be launched into a sustained decretion disk.

Yet another question is the origin of Be stars' rapid rotation.
One possibility is that Be stars are formed as isolated single stars. 
In this {\it single-star} path, Be stars may get their rotation from core contraction along the main sequence (MS) and/or simply have high birth spins, \cite[e.g.][]{Meynet-Maeder2000,Mathew2008MNRAS,Hastings2021}.  
Another possible scenario is that Be stars attain spin from binary (or higher order) interactions including binary mass transfer or stellar mergers \citep[][]{Kriz_Harmanec1975BAICz,Pols1991AA,Dallas2022ApJ,Shao_Li2014ApJ}. 
The binary channel has recently gained favor due to a multitude of observations: Gaia space satellite measurements of the relatively lower binary fraction of Be vs.\ B field stars \citep[][]{Dodd2023MNRAS}; identification of Be star binaries with sub-dwarf (sdO, stripped helium star) companions \citep[][]{El-Badry_Quataert2021MNRAS,Ramachandran2023AA,Wang2023AJ}; and the discovery and characterization of Be stars with \ac{CO} companions, referred to as Be X-ray binaries \citep[Be-XRBs;][]{Reig2011,Fortin2023AA}.

In this work we model Be-XRBs, a subclass of \acp{HMXB}\footnote{The others are supergiant XRBs \citep[][]{Corbet1986MNRAS} and supergiant Fast X-ray Transients \citep[][]{Negueruela2006ESASP} which primarily accrete through strong stellar winds from an evolved donor.} in which the \ac{CO} accretes material from the decretion disk emanating from the central Be star, producing variable X-ray emission correlated with the H$\alpha$ emission from the disk \citep{Okazaki1997AA,Reig2011}.
Nearly all Be-XRBs with characterized \ac{CO} companions harbor \acfp{NS} \citep{Reig2011}. 
There is one exception with a tentative BH Be-XRB claimed \citep[MWC-656;][however \citealp{Janssens2023AA} favor a \ac{NS} or lower mass companion]{Casares2014Nature}. 
The general picture views Be-XRBs as the descendants of Be+sdO binaries which remain bound after the sufficiently massive helium (He) star explodes in a \ac{SN}.
Be-XRBs have therefore garnered broad interest from binary stellar evolution studies as they encode the physics of \acfp{SN}, \acf{CE}, \acf{MT}, etc.\ into their resulting observed populations \citep[e.g.][and others]{Pols1991AA,Portegies-Zwart1995AA,vanBever-Vanbeveren1997AA,Belczynski-Ziolkowski2009,Shao_Li2014ApJ,Grudzinska2015MNRAS,Vinciguerra2020,Xing_Li2021ApJ,Misra2023AA,Liu2024MNRAS}.
The majority of the aforementioned studies use population-synthesis modeling techniques wherein single-star models \citep[SSE,][]{Hurley2000MNRAS.315..543H} and analytic prescriptions are combined to approximate the effects of \ac{MT} on the star's evolution, since computing self-consistent binary evolution sequences has hitherto been computationally infeasible.
Therefore, the SSE formalism cannot self-consistently model the the angular momentum evolution of stars as they interact and evolve, which is a critical formation scenario explaining Be stars in Be-XRBs.

To simultaneously model binary stellar evolution and mass transfer, we employ the next-generation BPS code \ac{posydon}. \ac{posydon} relies on a new modeling paradigm, wherein binary stellar tracks computed with \acl{mesa} \citep[\mesa{};][]{Paxton2011ApJS,Paxton2015ApJS,Paxton2013ApJS,Paxton2018ApJS,Paxton2019ApJS} are integrated within a BPS framework \citep{Fragos2023ApJS}.
This novel treatment includes the self-consistent evolution (solving the equations of stellar structure and composition) of individual stellar spins accounting for tides, mass transfer, and spin-orbit coupling.
We investigate the predicted rotational distribution of Be-XRB BPS models and compare our model Be-XRB population to the Galactic sample of observed Be-XRBs from \cite{Fortin2023AA}.

In \autoref{sec:observed_sample} we describe our data selection of Galactic Be-XRBs, and detail the relevant BPS modelling assumptions in \posydon{} and our adopted Be-XRB selection criteria in \autoref{sec:methods}.
We compare our models to data in \autoref{sec:results} and show how rotational modelling can naturally explain features of the observed Galactic Be-XRB population. 
Finally in \autoref{sec:discussion} and \autoref{sec:conclusions} we discuss our findings and summarize with concluding remarks.

\section{Observed Sample}
\label{sec:observed_sample}
For this study we use the newly updated catalog of 159 Galactic HMXBs compiled by \cite{Fortin2023AA}.
We select the subset of the catalog that includes systems with a firm Be-XRB classification and orbital solutions, excluding systems with tentative or hybrid classifications such as ``Be?" or ``Be/Sg".
We also exclude most gamma-ray binaries with Be companions since these systems may produce their high energy emission through interactions outside the canonical picture of a \ac{NS} accreting from the decretion disk \citep[e.g. white dwarf companions, magnetic field interactions, etc.,][]{Langer2020AA}.
Thus we include the systems which are seen to interact with the decretion disk of the donor \citep[namely PSR~B1259-63,][]{Hare2023ApJ}.
We consider two sub-samples of our selected Be-XRBs: $i$) all Be-XRBs with orbital period measurements, and $ii$) Be-XRBs with both orbital period and eccentricity measurements. Our resulting observed samples contain 46 and 18 systems respectively, listed in \autoref{table:Fortin_BeXRB_params}.

This sample contains binaries with Be donor stars that are preferentially early B-type (B0--B2), span a range of orbital periods from 10-1000 days, and are generally eccentric.
Their properties are listed in \autoref{table:Fortin_BeXRB_params}.
\section{Methods}
\label{sec:methods}

\subsection{Population synthesis physics}
\label{sec:methods:pop_synth_phys}
To simulate a Galactic Be-XRB population, we use the \posydon{} v1 \citep{Fragos2023ApJS} binary population synthesis (BPS) code to evolve a population of two million binaries at solar metallicity with $Z=0.0142$ and $Y=0.2703$ \citep{Asplund2009ARAA}.
In this section we provide a brief summary of our fiducial model parameters and key physical assumptions in our population synthesis.

To simulate a realistic population of interacting binaries, we draw their initial masses ($M_1, M_2$) and orbital periods ($P_\mathrm{orb}$) using observationally motivated distributions.
We draw primary masses using a Kroupa power law initial mass function \citep[IMF,][]{Kroupa2001MNRAS} with an index $\alpha=2.3$, with a minimum mass of $7~M_\odot$ and maximum mass of $120~M_\odot$, corresponding to the limits of the detailed binary grids in \posydon{} v1.
Secondary masses are determined by sampling uniformly in the mass ratio $q=M_1/M_2$ in the bounds $[0,1]$ \citep{Sana2013AA} with a minimum allowed mass of $1~M_\odot$. 
Initial orbital periods are drawn following \cite{Sana2013AA} in the range [$1$ day - $10^{3.5}$ day], and following a flat-in-log distribution from $1$ day down to $0.35$ day as described in \cite{Fragos2023ApJS}.
All binaries are initialized with circular orbits with stellar spins set to the orbital period, implying efficient tidal synchronization during the pre-main-sequence.

Binaries are evolved forward in time for 10 Gyr or until they end their life as a merger following an unsuccessful \ac{CE}, a double \ac{CO} merger, or a binary disruption. 
In the \posydon{} framework, phases of binary evolution where \ac{RLO} mass transfer can occur are modeled using the results of detailed binary evolution simulations with \mesa{}, where stars are evolved simultaneously with the orbit allowing self-consistent treatment of mass-transfer stability.
We use initial-final linear interpolation and $k$-nearest neighbor classification in all binary grids as described by \cite{Fragos2023ApJS}, to evolve a binary from an initial state through phases of evolution (e.g. \ac{RLO} \ac{MT} to \ac{SN}, \ac{CE}, etc.).
To simulate SNe and \ac{CE} phases, we utilize standard BPS prescriptions as described below.

Following \posydon{} v1, systems enter dynamically-unstable \ac{RLO} \ac{MT} if the \ac{MT} rate exceeds $0.1~M_\odot~\mathrm{yr}^{-1}$, or if the stellar model expands beyond the outer Lagrange point $\mathrm{L}_2$, triggering the onset of common envelope evolution.
We model \ac{CE} using the $\alpha_\mathrm{CE}-\lambda_\mathrm{CE}$ formalism \citep{Webbink1984ApJ,Livio_Soker1988ApJ}, where energy balance is used to determine if the envelope can be unbound with the orbital energy of the binary. 
The fraction of orbital energy released during the common envelope, $\alpha_\mathrm{CE}$, we set to $1$, and the binding energy parameter of the envelope, $\lambda_\mathrm{CE}$, is calculated directly from profiles of our detailed \mesa{} models using a core-envelope boundary defined by the mass shell where the central hydrogen (H) mass fraction drops below $10\%$.
Systems that go through a successful CE with two H-rich stars are subsequently evolved as a detached system (i.e. two single stars).

We model \ac{SN} explosions and compact object formation using the \cite{Patton-Sukhbold-2020MNRAS} prescription for core-collapse \ac{SN} (CCSNe), and the \cite{Podsiadlowski2004ApJ} prescription for electron-capture \ac{SN} (ECSNe) for stars with final He core masses in the range
of $1.4-2.5~M_\odot$.
NSs receive natal kicks during the SN explosion which we draw from a Maxwellian distribution with a dispersion of $\sigma_\mathrm{CCSNe}=265~\mathrm{km~s^{-1}}$ for CCSNe \citep{Hobbs2005MNRAS}, and a dispersion of $\sigma_\mathrm{ECSNe}=20~\mathrm{km~s^{-1}}$ for ECSN \citep{Giacobbo_Mapelli2019MNRAS}.
The maximum mass of \acp{NS} is taken as $2.5~M_\odot$ \citep{Oppenheimer1939PhRv}.

\subsubsection{Rotationally-limited accretion in detailed binary models}
\label{sec:methods:rot_limit_accretion}
\posydon{} v1 binary models initialize two hydrogen-rich stars on the main sequence (HMS-HMS grid) in circular orbits, with stellar spins synchronized to the orbital period.
We include self-consistent modeling of the angular momentum evolution of both stars including the effects of tides, spin-orbit coupling, mass-loss through stellar winds, and \ac{RLO} mass-transfer \citep[][]{Fragos2023ApJS}.

During HMS-HMS evolution we follow \cite{Langer1998AA} and implement rotationally-limited accretion, assuming fully conservative \ac{RLO} \ac{MT} but including an enhanced stellar wind from the accretor ($\dot M_\mathrm{w}$) which regulates the stellar rotation below critical:
\begin{equation}
    \dot M_\mathrm{w}(\omega) = \dot M_\mathrm{w}(0) \left(\frac{1}{1-\rot{}} \right)^{0.43},
    \label{eqn:rotation_enhanced_wind}
\end{equation}
where $\omega$ is the equatorial surface angular rotation rate of the star, $\dot{M}_\mathrm{w}(0)$ is the wind mass loss rate of the star with zero rotation, $\omega_\mathrm{crit}^2 = (1 - {L}/{L_\mathrm{Edd}})GM/R^3$ is the critical rotation angular frequency with the stellar luminosity $L$, equatorial radius $R$, and gravitational constant $G$. The $0.43$ exponent in \autoref{eqn:rotation_enhanced_wind} is taken from \cite{Langer1998AA}.
We define the Eddington Luminosity as $L_\mathrm{Edd} = 4 \pi G M c / \kappa$, where we assume the opacity, $\kappa$, is dominated by Thompson scattering $\kappa=0.2(1+X)~\mathrm{cm^2~g^{-1}}$, where $X$ is the hydrogen abundance of the star. 
Thus, the accretor may initially accept the mass transferred when its rotation is sub-critical, but at later times reject a significant amount of mass as the accretor approaches critical rotation, effectively halting further mass accretion.

\subsubsection{Rotational Evolution in the detached phase}
\label{sec:methods:rot_det_phase}
Binary systems may exit the two hydrogen rich stars (HMS-HMS) binary phase under multiple stopping criteria: the primary depleting central carbon (leading to a \ac{SN}) or the binary undergoing unstable mass transfer (leading to \ac{CE}).
Systems which either remain bound after the first SNe, or survive a \ac{CE} by successfully ejecting the envelope, will enter \textit{detached evolution} of a star+\ac{CO}. 

We match the secondary stellar model at the end of the HMS-HMS simulations to an equivalent precomputed and nonrotating single-star model (matching criteria are evolutionary-phase dependant as described in \citealp{Fragos2023ApJS}) such that the evolutionary state and mass between the two is similar.
To compute an initial angular rotation rate $\omega_i$ for our newly matched single-star model, we use the total rotational angular momentum $J_\mathrm{rot}$ and moment of inertia $I$ of the corresponding (rotating) primary or secondary from the end of its HMS-HMS evolution: $\omega_i = J_\mathrm{rot}/I$.
Using this initial rotation rate for our single star model, we continue evolving the orbit and stellar model until either \ac{RLO} occurs or the stellar model reaches the end of its life.
The orbital evolution equations include tidal evolution with spin-orbit coupling and angular momentum loss from stellar winds.
Our pre-computed single-star models are evolved without rotation and therefore cannot account for the back reaction onto the star's structure and evolution.

\subsection{Be-XRB modeling}
\label{sec:methods:Be_XRB_modeling}
For our study, we first select a \textit{baseline} XRB candidate population: all systems which $i)$ remain bound after the first \ac{SN}, and $ii)$ contain either a \ac{NS} or \ac{BH}. 
These systems will generally be eccentric with a stellar companion, and may exhibit detectable X-ray emission during their lifetime, depending on the orbital evolution and mass of the donor. 
It is this baseline XRB population that we use to extract subpopulations for comparisons to data.

Our modeling choices and selection criteria to categorize simulated binaries from the baseline population as Be-XRBs are described in the following subsections: donor rotation (\S\ref{sec:methods:donor_rotation}), donor mass (\S\ref{sec:methods:donor_mass}), disk interaction (\S\ref{sec:methods:disk_interaction}), and X-ray luminosity (\S\ref{sec:methods:Lx}).

\subsubsection{Stellar Rotation \& the Be phenomenon}
\label{sec:methods:donor_rotation}
We adopt a conservative lower limit of $\rot{} > 0.5$, above which donors are assumed to form a sustained decretion disk and appear as Be stars\footnote{Whether the disk itself is observable depends upon the strength of H$\alpha$ emission and the duty cycle of the disk. For Galactic Be-XRBs, a negative H$\alpha$ equivalent width is the threshold for detection \citep{Reig2011}, which, given our mass range and disk model, is achieved with modest disk sizes of 5-10 stellar radii \citep{Grundstrom_Gies_2006ApJ}, well within the radii of our adopted disk model.}.
While this lower limit on $\rot{}$ is conservative compared to the higher values often adopted in BPS Be-XRB studies \cite[e.g.][]{Misra2023AA,Liu2024MNRAS}{}, it is well within the range of recently reported measurements of the rotation rates of Be stars from as low as $\rot{}=0.3$, up to nearly critical values \citep{Fremat2005AA,Cranmer2005ApJ,Zorec2016AA,Balona-Ozuyar2021ApJ}.

The diversity of measured rotation rates may be a consequence of the intrinsic efficiency of the Be phenomenon as a function of rotation \citep{Cranmer2005ApJ}, evolutionary state \citep{Cochetti2020AA}, metallicity, etc. and or systematic biases from inferring rotation rates of rapidly rotating stars \citep{Townsend2004MNRAS,Fremat2005AA}.
To explore this further, we also compare different regions of the rotational distribution within our threshold cutoff of $\rot{} > 0.5$, to see if their origins can be interpreted through stellar and binary evolution processes.

\subsubsection{Stellar mass}
\label{sec:methods:donor_mass}
We chose a conservative mass cutoff for the donor star's mass $M_\mathrm{d} \gtrsim 3~M_\odot$, to investigate the role of rotation on the resulting Be-XRB subpopulation.
While early O-type stars are not observed to exhibit the Be phenomenon, we refrain from imposing an upper mass limit in our definition of Be-XRBs as we seek to reproduce this feature self-consistently.

This lower limit on $M_\mathrm{d}$ aligns with other BPS Be-XRB studies which range from as low as $\simeq 3~M_\odot$ \citep[e.g.][]{Xing_Li2021ApJ,Vinciguerra2020} to be consistent with the spectral definition of B stars \citep{Harmanec1988BAICz}, or as high as $\simeq 6~M_\odot$ \citep[e.g.][]{Liu2024MNRAS,Misra2023AA}, based on masses most often reported in Be-XRB catalogues.

\subsection{Accounting for observable lifetime}
\label{sec:methods:SFH}
To compare our BPS models to the Galactic sample of Be-XRBs, we apply a constant star formation history to our population at solar metallicity.
We take $500$ random, uniformly distributed time samples for the evolution of systems during their detached phase, where all samples for an individual system are weighted by the lifetime of the detached phase:
\begin{equation}
    w_{t_\mathrm{det}} = \frac{ t_\mathrm{det} }{ t_\mathrm{H} },
    \label{eqn:w_t_det}
\end{equation}
where the detached phase lifetime, $t_\mathrm{det}$, is defined as the time between the first \ac{SN} and ends if \ac{RLO} \ac{MT} is initiated or if the stellar companion reaches the end of its life, and an arbitrary scaling factor $t_\mathrm{H}$ is taken as the Hubble time.
This weighting reflects the fact that longer-lived systems should appear more frequently for an impulsive, unbiased observation of the Galaxy.
In total, from our initial population of 2 million binaries, our baseline \ac{CO}+star population contains $1.5 \times 10^{8} $ unique evolutionary samples for the time evolving parameters of detached binaries. 

\subsubsection{Disk Interaction}
\label{sec:methods:disk_interaction}
To capture the transient X-ray emission of Be-XRBs, we adopt a decretion disk model to select systems which are in orbits close enough to interact and accrete material from the disk.
We follow the formalism from \cite{Liu2024MNRAS}, assuming an isothermal temperature profile with $T=0.6~T_\mathrm{eff,d}$ \citep{Carciofi_Bjorkman2006ApJ}, where $T_\mathrm{eff,d}$ is the effective temperature of the donor.
The maximum unperturbed disk radius $r_\mathrm{disk}$ is taken as the sonic point $r_\mathrm{sonic}$ where the azimuthal velocity of the disk transitions from supersonic to subsonic \citep{Krticka2011AA}:
\begin{equation} 
    r_\mathrm{sonic} = 0.3~(v_\mathrm{Kep}/c_s)^2 R_\mathrm{d},
    \label{eqn:r_sonic}
\end{equation}
where the equatorial Keplerian orbital velocity $v_\mathrm{Kep} = \sqrt{G M_\mathrm{d}/R_\mathrm{d}}$, and the sound speed $c_s = \sqrt{2 \mathrm{k}_\mathrm{B} T_\mathrm{eff,d} / m_\mathrm{p}}$, where $\mathrm{k}_\mathrm{B}$ is Boltzmann's constant and $m_\mathrm{p}$ is the proton mass. 
Beyond the sonic point, the Keplerian orbital velocity in the disk is subsonic, leading to weak viscous coupling.

As there exists a diversity of choices among disk models between BPS studies, we also consider a simpler treatment implemented by \cite{Misra2023AA}, where the maximum disk radius is taken as $100~R_\mathrm{d}$, motivated by measurements of Be disk radii ranging from $30-150~R_\mathrm{d}$ \citep{Klement2017AA}.
We compare both models further in \S\ref{sec:results:disk_model_comp}.

We define binaries which have periapse distances, $r_\mathrm{peri} = a(1-e)$ where $a$ is the orbital semi-major axis, less than the disk radius, $r_\mathrm{peri} < r_\mathrm{disk}$, as interacting sources, where the X-ray luminosity is calculated in \S\ref{sec:methods:Lx}.
We assume the disk has a constant radius as long as the donor is above the threshold rotation rate for the Be phenomenon to occur.

\begin{figure*}[ht]
    \centering
    \includegraphics[width=2\columnwidth]{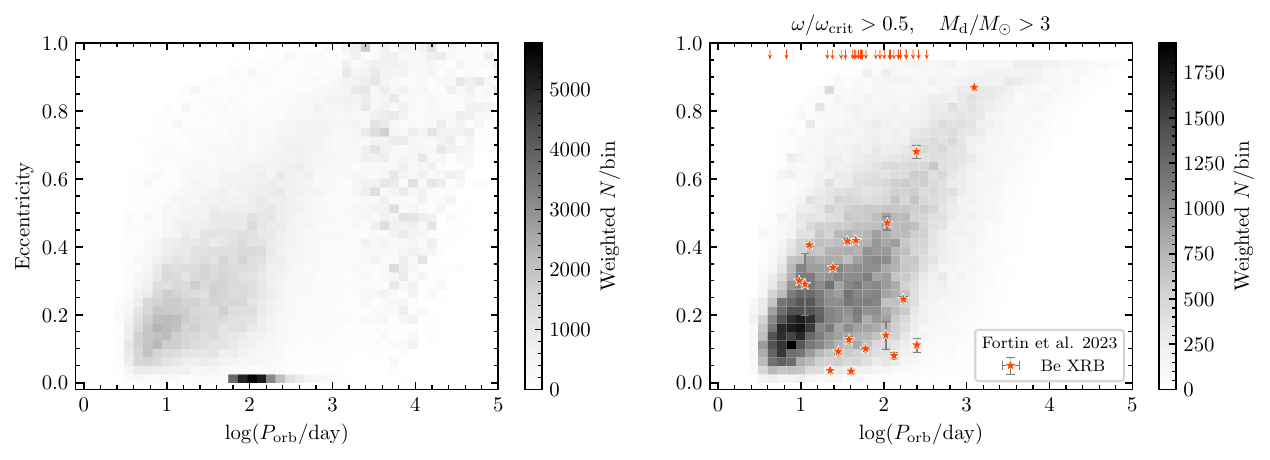}
    \caption{Two-dimensional histogram of the time-sampled orbital properties of our binary population, weighted by their detached lifetime $w_{t_\mathrm{det}}$. The left panel shows the baseline population (a non-degenerate star with a \ac{CO} companion) of candidate XRBs while the right panel shows our model Be-XRB subpopulation with our selected sample of Galactic Be-XRBs (from \cite{Fortin2023AA}, listed in \autoref{table:Fortin_BeXRB_params}) is overlaid for comparison.
    Orange stars and arrows indicate observed Be-XRBs with $P_\mathrm{orb}-e$ and only $P_\mathrm{orb}$ measurements, respectively. 
    }
    \label{fig:HMXB_population_p_vs_e_data}
\end{figure*}

\subsubsection{X-ray Luminosity}
\label{sec:methods:Lx}
To simulate the X-ray luminosity of the Be-XRBs in our sample and determine whether a given system is observable, we adopt the formalism presented by \cite{Liu2024MNRAS}. 
The X-ray luminosity is calculated by first deriving a scaling relation between the mass accretion rate onto the CO ($\dot M_\mathrm{CO}$) as a function of the periastron separation ($r_\mathrm{peri}$) and base gas surface density ($\Sigma_0$) of the disk, $\dot M_\mathrm{CO} \propto r_\mathrm{peri}^{-2}~\Sigma_0$, using the smoothed-particle hydrodynamics (SPH) simulations by \cite{Brown2019MNRAS}.
Then \cite{Liu2024MNRAS} normalize the accretion rate by deriving another scaling relation between decretion disk surface densities and the mass of the donor star ($M_\mathrm{d}$), $\Sigma_0 \propto M_\mathrm{d}^{1.44}$, using observations of Milky Way Be star disks from \cite{Vieira2017MNRAS}.
Finally they adjust correction factors accounting for uncertainties in the SPH simulations and conversion between bolometric and X-ray luminosity, to best match their synthetic population to the empirically-derived peak X-ray luminosity function from \cite{Raguzova_Popov2005AA}:
\begin{eqnarray} 
     \log( L_\mathrm{X} / [10^{35} \mathrm{~erg~s^{-1}}] ) = \quad &  \nonumber{} \\ 
     4.53 \pm 0.66 - (1.50 &\pm 0.33)  \log(P_\mathrm{orb}/\mathrm{d}). 
     \label{eqn:Raguzova_Popov2005}
\end{eqnarray}

The final equations for calculating the X-ray luminosity for our binaries are as follows:
\begin{eqnarray}
    &L_\mathrm{bol} = \epsilon \dot M_\mathrm{CO} c^2 = f_\mathrm{corr} \times 5 \times 10^{36} ~\mathrm{erg~s^{-1}} \nonumber{}\\
    &\times \left(\frac{\epsilon}{0.2}\right) \left[\frac{(1-e)a}{100~R_\odot}\right]^{-2}
    \left(\frac{\Sigma_0}{0.015 ~\mathrm{g~cm^{-2}}}\right) \left(\frac{M_\mathrm{CO}}{1.4~M_\odot}\right)^{2},
    \label{eqn:bolometric_luminoisty}
\end{eqnarray}
where $f_\mathrm{corr}=0.5$ is the correction factor between the bolometric and X-ray luminosity, $\epsilon=0.2$ is the efficiency of converting gravitational potential energy to radiation, $M_\mathrm{CO}$ is the compact object mass, and $\Sigma_0$ is the base gas density of the donor star's decretion disk:
\begin{eqnarray}
    \log( \Sigma_0 / [\mathrm{g~cm^{-2}}] ) = & \nonumber{}\\  1.44 \log( M_\mathrm{d} &/ M_\odot )  - 2.37 + \mathcal{N}(0, 0.52), \label{eqn:base_gas_density}
\end{eqnarray}
where $M_\mathrm{d}$ is the mass of the donor star and $\mathcal{N}(0, 0.52)$ is a normal distribution with zero mean and standard deviation of $0.52$, encapsulating the intrinsic scatter seen in the observed distribution of Be disk surface densities \citep{Vieira2017MNRAS}. For a detailed derivation of \autoref{eqn:bolometric_luminoisty} and \autoref{eqn:base_gas_density}, see \cite{Liu2024MNRAS}.

To model the detection of high energy emission of Galactic Be-XRBs, we use a threshold X-ray luminosity of $L_X = 10^{34}~\mathrm{erg~s^{-1}}$ above which binaries are considered detectable.
This limit is motivated by the distribution of X-ray luminosities for Be-XRBs observed in the Milky Way \citep[][]{Brown2018MNRAS}.

\section{Results}
\label{sec:results}

\subsection{Comparison to observed systems}
\label{sec:results:comp_to_obs}
To consider the accuracy of our Be-XRB modeling, we compare to observed distributions of systems in the plane of $P_\mathrm{orb}$ and eccentricity $e$ shown in \autoref{fig:HMXB_population_p_vs_e_data}, where all detached systems' time-sampled evolution is weighted by their detached lifetime as in \autoref{eqn:w_t_det}, such that the color scale shows the weighted number of binaries per bin given by: $\sum_{i \in \mathrm{bin}} w^{i}_{t_\mathrm{det}}$.
The left panel shows the baseline population containing CO+star binaries, with three main groups: an excess of circular systems at $\log(P_\mathrm{orb}/\mathrm{d}) \simeq 2$ due to a buildup of initially eccentric systems circularized by tides, moderately eccentric binaries at $\log(P_\mathrm{orb}/\mathrm{d})  \simeq 0.5-2.5$ from natal kicks during the SNe, and higher eccentricity, wide systems with $\log(P_\mathrm{orb}/\mathrm{d}) \gtrsim 3$ which are mostly low-mass stars with minimal interaction pre-SNe and are long lived due to their wide orbits.
While the group of circularized binaries contain MS donor stars, many binaries circularize during post-MS evolution, explaining the peak at larger orbital periods as these systems have longer detached lifetimes (i.e. live longer before initiating \ac{RLO} \ac{MT}).
The majority of binaries in our baseline population have evolved through stable \ac{RLO} \ac{MT}, with only $\sim 0.05\%$ having formed through a successful \ac{CE} event.

The right panel of \autoref{fig:HMXB_population_p_vs_e_data} shows the Be-XRB subpopulation after applying our modeling selection criteria onto the baseline population, with observed Be-XRBs from \cite{Fortin2023AA} overlaid (systems listed in \autoref{table:Fortin_BeXRB_params}).
Our selection criteria for Be-XRBs (\S\ref{sec:methods:Be_XRB_modeling}), are systems which simultaneously satisfy the following: rapid donor rotation ($\rot{}>0.5$), B-type donor mass ($M_\mathrm{d}>3~M_\odot$), interaction with the decretion disk ($r_\mathrm{peri} < r_\mathrm{disk}$), and X-ray bright ($L_X>10^{34}~\mathrm{erg~s^{-1}}$). 
As an initial comparison, our fiducial modeling can reproduce the broad orbital characteristics of Galactic Be-XRBs, with a high density of our models in the short period $\log(P_\mathrm{orb}/\mathrm{d})  \simeq 1$, low eccentricity $e \lesssim 0.2$ region, which we discuss further in \ref{sec:discussion:comp_to_obs}. 
Furthermore, our simulated population exhibits a similar correlation between orbital period and eccentricity to that in the observed population.

\begin{figure}[h]
    \centering
    \includegraphics[width=\columnwidth]{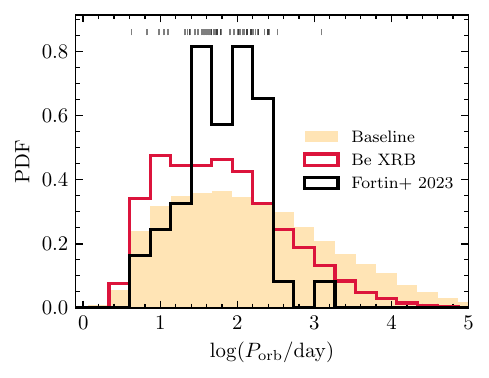}
    \caption{One-dimensional distributions of $\log_{10}(P_\mathrm{orb}/\mathrm{d})$ for our baseline (tan) and Be-XRB population (red) weighted by their detached lifetime ($w_{t_\mathrm{det}}$), compared to the observed period distribution (black) compiled from \cite{Fortin2023AA}. The small bars at the top of the plot indicate the individual systems within the histogram. }
    \label{fig:period_dist_Fortin}
\end{figure}

We also compare the one-dimensional $\log_{10}(P_\mathrm{orb}/\mathrm{d})$ distribution between our baseline and Be-XRB model populations in \autoref{fig:period_dist_Fortin} with the observed period distribution with 46 systems listed in \autoref{table:Fortin_BeXRB_params}.
Our models have a slight preference towards shorter orbital periods, with a flat peak around $\log(P_\mathrm{orb}/\mathrm{d}) \simeq 1-2$, compared to the observed population which appears to be peaking closer to $\log(P_\mathrm{orb}/\mathrm{d}) \simeq 2$.
Overall, the majority of the orbital period distribution of our Be-XRB models and the observed population overlap with minima and maxima around  $\log_{10}(P_\mathrm{orb}/\mathrm{d}) \simeq 0.5$ and $\log_{10}(P_\mathrm{orb}/\mathrm{d}) \simeq 3.5$, respectively.

\begin{figure}[h]
    \centering
    \includegraphics[width=\columnwidth]{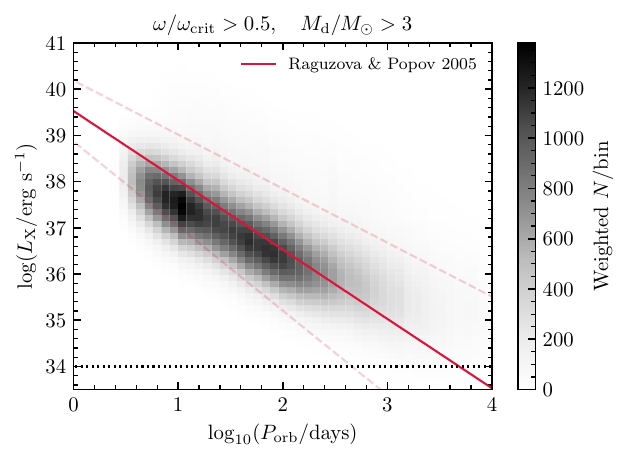}
    \caption{Predicted X-ray luminosity as a function of $P_\mathrm{orb}$ for our Be-XRB population weighted by their detached lifetime ($w_{t_\mathrm{det}}$). The red lines show the mean (solid) and uncertainty (dashed) from the empirical relation from \citet{Raguzova_Popov2005AA} provided in \autoref{eqn:Raguzova_Popov2005}. The black dotted line shows our chosen threshold X-ray luminosity.}
    \label{fig:Lx_v_period}
\end{figure}

We show in \autoref{fig:Lx_v_period} the simulated peak X-ray luminosities for our Be-XRB models before applying our X-ray luminosity threshold cutoff ($10^{34}~\mathrm{erg~s^{-1}}$ in \S\ref{sec:methods:Lx}), weighted by their detached lifetime.
We recover the empirical relation of \cite{Raguzova_Popov2005AA} in \autoref{eqn:Raguzova_Popov2005} showing a correlation between the orbital period and peak X-ray luminosity of Be-XRBs, where the bulk of our models are well above our chosen luminosity threshold.
Similar to \cite{Liu2024MNRAS}, we find a small population of ultra-luminous X-ray sources (ULXs) with $\mathrm{L_X} > 10^{39} ~\mathrm{erg~s^{-1}}$ and $\log(P_\mathrm{orb}/\mathrm{d}) \simeq 1$, which are faint in the figure due to their relatively rarity and shorter lifetimes.

\subsection{Effect of the Decretion Disk Model}
\label{sec:results:disk_model_comp}
We find that the choice of decretion disk model as described in \S\ref{sec:methods:disk_interaction} has a negligible effect on the simulated Be-XRB population.
In both disk models, we find that the majority of our binary models satisfy our disk interaction criteria, $r_\mathrm{peri} < r_\mathrm{disk}$.
Comparing the two maximum disk radii together for all donors we find $97\%$ of systems have $r_\mathrm{sonic} > 100 ~R_\mathrm{d}$.
When applying the disk models within the orbital configurations of our baseline population, we find  $97\%$  and $86\%$ of systems satisfy our interaction criterion using $r_\mathrm{sonic}$  and $100~R_\mathrm{d}$ respectively.
Furthermore,  \autoref{fig:comparing_disk_models} shows negligible differences in the rotation rates, masses, orbital parameters, and luminosities of Be-XRBs between the two disk models, suggesting the disk interaction criterion is not the leading uncertainty in the present study.

\subsection{Effect of the Rotation and Mass Cuts}
\label{sec:results:rot_mass_cuts}
\begin{figure*}[!ht]
    \centering
    \includegraphics[width=2\columnwidth]{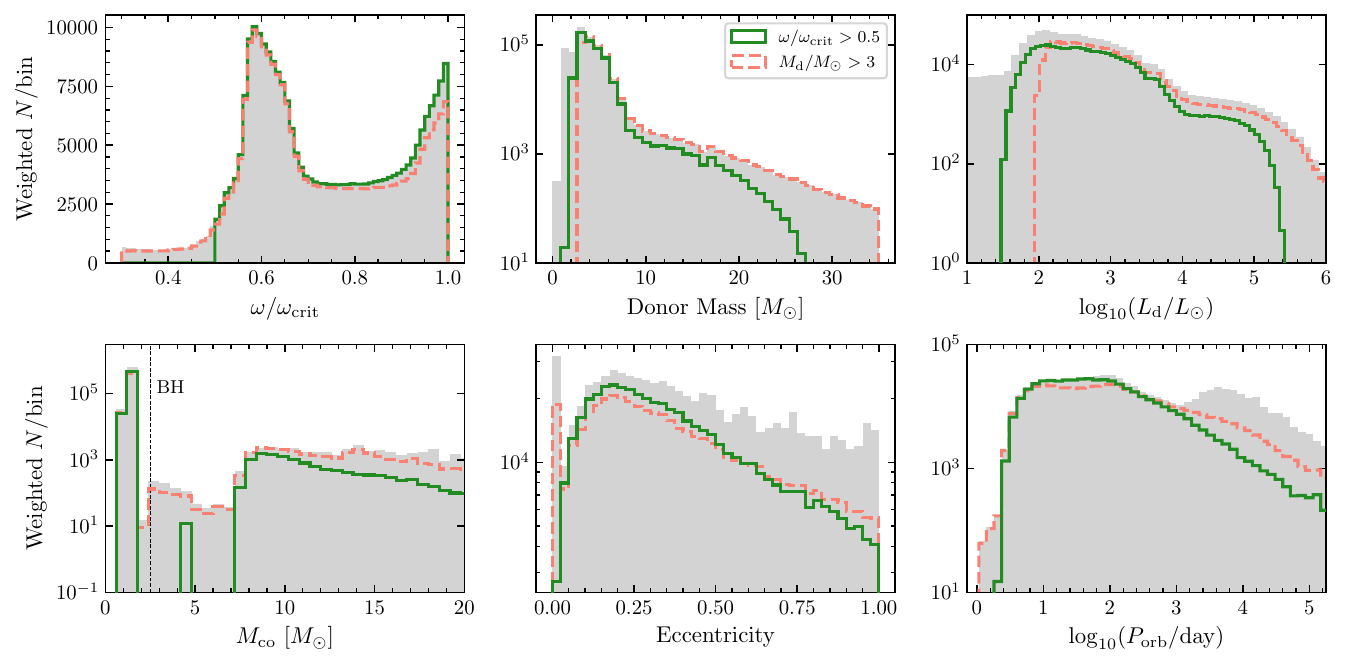}
    \caption{One-dimensional parameter histograms of the baseline population (gray), compared to cuts on rotation (solid green) or donor mass (dashed pink) alone.
    During the detached phase, we randomly sampled uniformly each binary's time evolution and weight samples by the lifetime of the binary ($w_{t_\mathrm{det}}$ defined in \autoref{eqn:w_t_det}). While we combine multiple selection criteria simultaneously to determine Be-XRBs in our population synthesis, we use these individual cuts to demonstrate the physical importance of each cut when constraining our Be-XRB population with rotational modeling ($\rot > 0.5$). 
    }
    \label{fig:individual_param_hist}
\end{figure*}

To interpret the physical effects of our Be-XRB selection criteria, we compare the critical rotation and donor mass constraints individually in \autoref{fig:individual_param_hist}, showing one-dimensional parameter distributions weighted by a system's detached lifetime from \autoref{eqn:w_t_det}.
This serves as an approximate comparison between traditional BPS selection criteria identifying Be-XRB donors using only mass cuts, and our criteria with \posydon{} which also rely upon self-consistent modeling of the rotation and accretion during the prior \ac{RLO} \ac{MT} phase.
The gray distribution shows the baseline population of CO+star binaries, while the solid green and dashed pink distributions show systems containing rapidly rotating donors with $\rot{} > 0.5$ or massive B-type donors with $M_\mathrm{d} > 3 ~M_\odot$, respectively.

Using a rotation criterion of $\rot{} > 0.5$ selects binary systems with donor masses in the range ${\sim}2-25 ~M_\odot$, more NS companions than BHs, and non-circular binaries peaking with $e \simeq 0.25$ and orbital periods around $\log_{10}(P_\mathrm{orb}/\mathrm{d})\sim 1-2$ with a tail falling off at wider orbital periods.
Cutting only on donor mass spans a wide range of rotation rates extending below our fiducial cut of $\rot{} > 0.5$, and including a large fraction of the highly spinning systems.
This cut also contains binaries with more BHs and selects a large fraction of circular systems.

The subpopulation identified as Be-XRBs using the critical rotation criterion alone is in better agreement with observations than using donor mass alone, which we discuss further in \secref{sec:discussion:rot_and_mass_cuts}.

\begin{figure}[!h]
    \centering
    \includegraphics[width=\columnwidth]{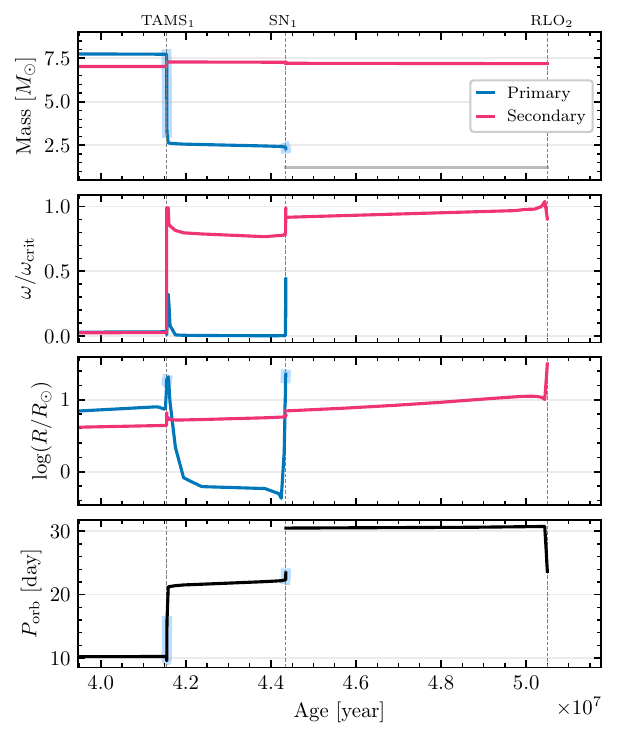}
    \caption{Characteristic time-series evolution of a Be-XRB from our population, evolving from ZAMS, through Case B \ac{MT} (light blue) until the first \ac{SN} forms a \ac{NS}, and into the detached phase. The discontinuity after SN$_1$ (e.g. total mass and radius) is from two effects: $i$) the detailed binary model used for HMS-HMS evolution is the nearest neighbor track (while IF interpolation was used to evolve the population), and $ii)$ the matching criteria taking the secondary star post HMS-HMS evolution mapping to an equivalent non-rotating single star model. This binary ends up merging during common envelope evolution after \ac{MT} is initiated by the secondary (RLO$_2$). The majority of Be-XRBs in our population eventually end through mergers since their mass ratio is extreme and are therefore more likely to enter dynamically unstable \ac{MT}.
    }
    \label{fig:characteristic_BeXRB_evolution}
\end{figure}

\subsection{Characteristic Evolution of Be-XRBs}
\label{sec:results:characteristic_be_XRB_evol}
It has long been suggested that Be-XRBs primarily form through the stable \ac{MT} channel, wherein the initially more massive star spins up the secondary through MT while on the MS, before subsequently experiencing core-collapse, leaving behind a compact object with a Be companion in a Be-XRB \citep[][]{Rappaport_vdHeuvel1982IAUS,Pols1991AA,Portegies-Zwart1995AA,deMink2013ApJ}.
In this work we confirm this standard formation scenario, finding that the vast majority of systems we categorize as Be-XRBs have gone through stable mass transfer while on the MS ($\gtrsim 99\%$).
A much smaller fraction of systems have not gone through any \ac{RLO} \ac{MT}, and a negligible fraction ($\lesssim 0.01\%$) are forming through a common envelope.
While more binaries in our initial population generally survive common envelope, they often do not become Be-XRBs as the surviving helium cores are not massive enough ($\lesssim 2.5~M_\odot$) to undergo a \ac{SN}, forming a \ac{CO}.

Branching fractions for the baseline and Be-XRB populations are presented in \autoref{table:mass-transfer-cases}, categorized by a system's \ac{MT} history during their MS evolution.
To summarize the details of a system's evolution which may involve multiple phases of \ac{RLO} \ac{MT}, we use standard definitions of \ac{MT} phases: Case A, Case B, and Case C, where each letter corresponds to \ac{MT} during the MS, Hertzsprung-gap (or stripped He star), or an evolved state, respectively \citep{Iben1991ApJS}.
Systems which undergo successive phases of \ac{MT} are denoted with multiple letters separated by slashes (e.g. we denote Case A followed by Case B as Case A/B) as introduced by \cite{Fragos2023ApJS}. 

In the baseline population of CO+star binaries, the largest number of systems go through Case B ($34\%$), Case B/C/BB MT ($20\%$), or have no \ac{RLO} \ac{MT} on the MS ($15\%$).
Case B/C/BB \ac{MT} describes stars which initiate \ac{MT} after reaching \ac{TAMS} continuing into the star's giant phase, followed by another phase of \ac{MT} from the now stripped He star as it leaves the He MS.
While Case B and Case B/C/BB systems have a high probability of evolving into Be-XRBs ($56\%$ and $87\%$ respectively), systems which do not enter \ac{RLO} have a low probability ($2\%$) of becoming Be-XRBs, making up $\lesssim 0.1\%$ of the entire weighted Be-XRB population.
We find almost half our weighted Be-XRB population ($47\%$) have gone through Case B/C/BB evolution, with the remaining majority going through Case B ($20\%$), Case B/BB ($15\%$), or Case B/C ($14\%$).
In total, ${\simeq}64\%$ of the weighted Be-XRBs have gone through a phase of Case BB MT, while those systems only going through Case A or Case A/B contribute ${\simeq}1.5\%$.

To further investigate the physical processes leading to the Be-XRB phase identified from \autoref{sec:results:rot_mass_cuts}, we show in \autoref{fig:characteristic_BeXRB_evolution} the time-series evolution of one such system taken at random from our population synthesis models.
This binary goes through Case B \ac{MT} initiated by the primary with an initial mass of ${\sim}7.5~M_\odot$, where about $5~M_\odot$ of material is transferred through \ac{RLO} to the secondary star, leaving behind the ${\sim}2.5~M_\odot$ He core.
During \ac{MT}, the secondary quickly reaches critical rotation from the mass accretion \citep{Packet1981AA}, causing the majority of the subsequent mass transferred from the primary to be lost through the rotationally-enhanced winds of the accretor (\autoref{eqn:rotation_enhanced_wind} described in \ref{sec:methods:rot_limit_accretion}).
As a result of the mass and angular momentum loss from the system, the binary widens, increasing the orbital period which stabilizes the system as it naturally exits the \ac{MT} phase into a detached binary containing a He and H-rich star.
Then  $\sim 3~\mathrm{Myr}$ after \ac{MT}, the primary, which is now a stripped He star, undergoes a \ac{SN} forming a \ac{NS} in an $e \simeq 0.18$, $30$ day orbit with a $7.4~M_\odot$ rapidly rotating companion.

This marks the beginning of this binary's Be-XRB phase, where the secondary has evolved ${\simeq}20\%$ of its MS lifetime.  
As the secondary continues on the MS, its radius expands, causing a steady increase in $\rot{}$ until it reaches \ac{TAMS}.
While the eccentricity remains nearly unchanged during the detached phase, there is a short period when the eccentricity increases as the stellar spin angular momentum is redistributed by tides into the orbit.
If the spin angular momentum of the star is comparable to the orbital angular momentum, this can lead to eccentricity increases of nearly $e \simeq 0.1$ in the most extreme cases, and in general extends the lifetime of binaries in their eccentric configuration.

\begin{figure}[!h]
    \centering
    \includegraphics[width=\columnwidth]{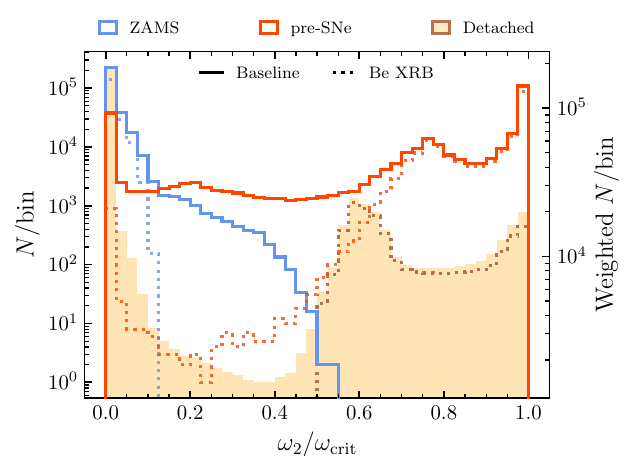}
    \caption{Rotational evolution of our baseline and Be-XRB population at different phases of stellar evolution: at ZAMS, pre-SN, and the subsequent time-sampled evolution during the detached XRB phase, weighted by the system's lifetime ($w_{t_\mathrm{det}}$). Both ZAMS and pre-SN distributions contain instantaneous values for an individual system. The solid lines show the baseline population comprised of CO+star binaries, while the dotted lines show the Be-XRB population. The transformation between the ZAMS rotational distribution at relatively low values of $\rot{}$, and the pre-SN distribution is a direct result of the angular momentum gained through \ac{RLO} \ac{MT} during their HMS-HMS evolution.
    }
    \label{fig:omega_evolution}
\end{figure}

Since our present study is focused on the role of rotation in Be-XRB modeling, we show in \autoref{fig:omega_evolution} the evolution of the rotational distribution of secondary stars at multiple evolutionary phases: at ZAMS, the moment before the first SNe (post \ac{MT}), and a time sampled distribution during the detached phase, weighted by $w_{t_\mathrm{det}}$ from \autoref{eqn:w_t_det}.
The dashed lines tracing the Be-XRB subpopulation, show that all systems have relatively low rotation rates at ZAMS with $\rot{} \lesssim 0.2$, and are spun up through \ac{RLO} \ac{MT}.
Prior to the first \ac{SN}, the rotational distribution has shifted closer to critical with the majority of systems having  $\rot{} \gtrsim 0.5$.
Then, as systems evolve during the detached phase, depending on their orbital and donor properties, tides serve to synchronize the rotation rate with the orbital period, effectively slowing down the donor stars, causing the peak of the distribution to shift to lower values at $\rot{} \simeq 0.6$.
Also present in the detached rotational distribution is a secondary peak at $\rot{} \simeq 0.99$, caused by a star's gradual expansion as it approaches \ac{TAMS}, decreasing $\omega_\mathrm{crit}$ \citep[][]{Meynet-Maeder2000,Hastings2021}.
We also note the large bifurcation between slowly rotating ($\rot{} \simeq 0$) and moderate to rapidly rotating systems ($\rot{} \gtrsim 0.5$), due to the strong dependence of tidal evolution on the separation of the binary and radius of the star \citep{Hut1981A&A}.

As a result of \ac{MT} during the HMS-HMS evolution, secondaries are spun up to a wide distribution of rotation rates before entering their detached evolutionary phase with peaks at $\rot{} \simeq 0$, $\rot{} \simeq 0.75$, and $\rot{} \simeq 0.99$.

\begin{figure}[!h]
    \centering
    \includegraphics[width=\columnwidth]{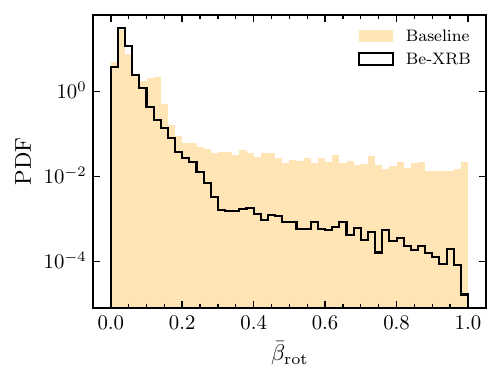}
    \caption{Rotationally-limited mass-transfer efficiency ($\bar{\beta}_\mathrm{rot}$ as in \autoref{eqn:mean_beta_rot}) of our population synthesis models evolving through their HMS-HMS evolution, weighted by their detached lifetime $w_{t_\mathrm{det}}$. In our binary models, the accretor can reject significant amounts of the matter transferred during \ac{RLO} \ac{MT} via its enhanced wind (as a function $\rot{}$), which keeps the star rotating below critical.}
    \label{fig:implicit_beta_distr}
\end{figure}

\begin{figure*}[t]
    \centering
    \includegraphics[width=2\columnwidth]{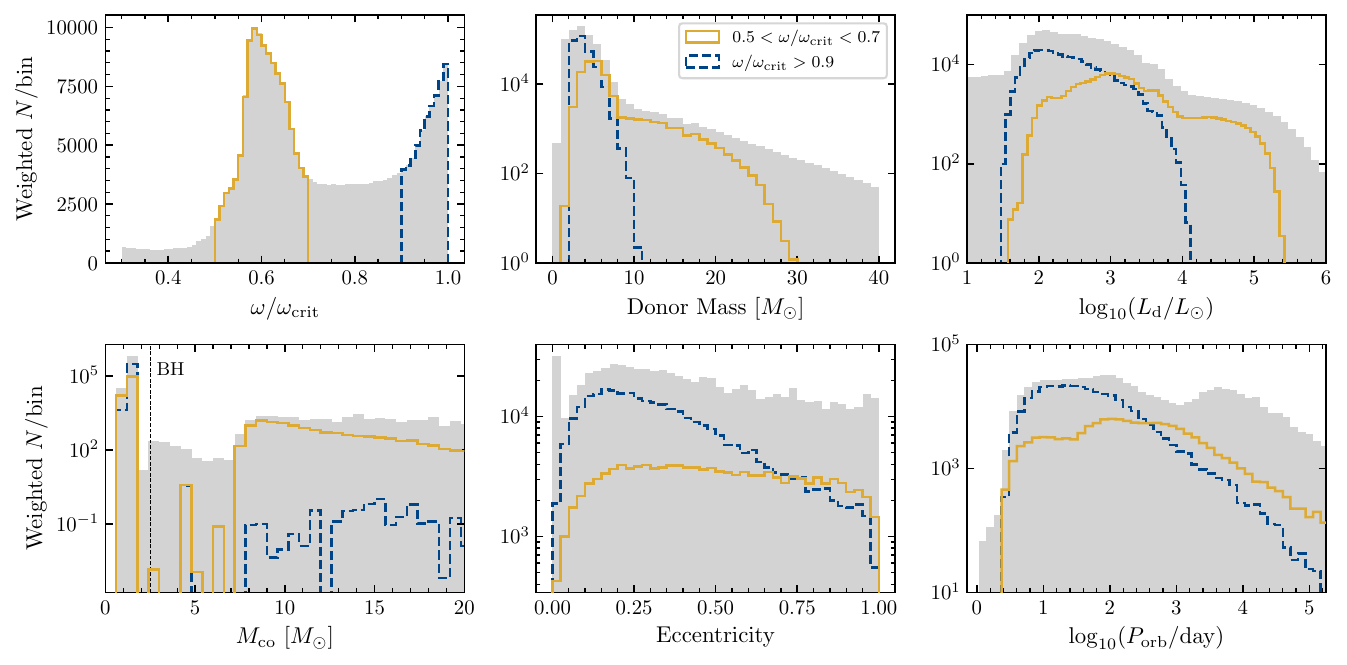}
    \caption{One-dimensional distributions of our time-sampled detached binary population (weighted by their detached lifetime $w_{t_\mathrm{det}}$), showing the intermediate ($\rot{} \simeq 0.6$, solid yellow) and critical ($\rot{} \simeq 0.99$, dashed blue) rotation peaks in donor and CO mass, and orbital properties. Overall the two rotational peaks are well separated in donor mass and luminosity, CO mass, and binary orbital properties.}
    \label{fig:bimodal_rot_hist}
\end{figure*}

\subsection{Mass Transfer Efficiency}
\label{sec:results:MT_efficiency}
A commonly explored parameter in BPS studies is the \ac{MT} efficiency, $\beta \equiv \Delta M_\mathrm{acc} / \Delta M_\mathrm{donor}$, determining how much mass transferred through \ac{RLO} is successfully accreted by the donor \citep{Hurley2002MNRAS}.
In order to compare our rotationally-limited accretion models described in \S\ref{sec:methods:rot_limit_accretion} to the standard population synthesis MT efficiency $\beta$, we calculate an effective rotationally-limited MT efficiency metric:
\begin{equation}
    \beta_\mathrm{rot} = \frac{ \Delta M_\mathrm{acc} }{ \Delta{M}_\mathrm{d,RLO} },
    \label{eqn:beta_rot_i}
\end{equation}
where $\Delta M_\mathrm{acc}$ is the mass gained by the accretor, and $\Delta{M}_\mathrm{d,RLO}$ is mass lost from the donor through a phase of Roche-lobe overflow \ac{MT} (defined as the sequences when the \ac{MT} rate exceeds a threshold value: $\dot M_\mathrm{RLO} > 10^{-15}~M_\odot ~ \mathrm{yr}^{-1}$).
Our models often contain multiple discrete phases of MT (e.g. case A/B/BB), which would produce multiple values for $\beta_\mathrm{rot}$ for an individual simulation.
Therefore, for this initial comparison, we perform a mass-weighted average of simulations with $n$ total MT phases: 
\begin{equation}
    \bar{\beta}_\mathrm{rot} = \frac{1}{M_\mathrm{tot}} \sum^{n}_{i=1} \beta_{\mathrm{rot},i} \cdot \Delta M_{\mathrm{acc},i},
    \label{eqn:mean_beta_rot}
\end{equation}
where $i$ indexes individual phases of MT, the mass weight $\Delta M_{\mathrm{acc},i}$ is the mass gained by the accretor during the $i$'th phase, and $M_\mathrm{tot}=\sum \Delta M_{\mathrm{acc},i}$ is the total mass gained by the accretor through the simulation.

This rotationally-limited MT efficiency $\bar{\beta}_\mathrm{rot}$ is then calculated for every model in our HMS-HMS grid and included in the same Initial-Final interpolation scheme used to evolve our binary population as described in \cite{Fragos2023ApJS}, which is then applied on our astrophysical binary population shown in \autoref{fig:implicit_beta_distr}. 

The baseline and Be-XRB populations have a strong peak at $\bar{\beta}_\mathrm{rot} \simeq 0.05$ with a long tail to higher \ac{MT} efficiencies.
The Be-XRB population is less likely to form through more conservative \ac{RLO} \ac{MT}  ($\bar{\beta}_\mathrm{rot} \gtrsim 0.2$) because $i$) they are intrinsically rare in the baseline as they are near regions of dynamically unstable \ac{MT}, and $ii$) the accretors are not spun up to large values of $\rot{}$, and are therefore short lived in their Be phase.

\subsection{Changing the critical rotation threshold for the Be phenomenon}
\label{sec:results:changing_omega_crit}
In \autoref{fig:bimodal_rot_hist}, we investigate the differences between simulated systems which population different peaks in the simulated rotational distribution. We show one-dimensional weighted distributions of our population during their detached phase, selecting systems with intermediate rotation at $\rot{} \simeq 0.6$ and systems with nearly critical rotational $\rot{} \simeq 0.99$.
Similar to the analysis in \autoref{fig:individual_param_hist}, we find the peaks in the bimodal rotational distribution are well separated in other binary and stellar parameters.
For the intermediate rotation population, we see that donors can be as massive as ${\sim} 30~M_\odot$, systems contain more BHs as CO companions, with a nearly flat eccentricity distribution, and an orbital period distribution peaking closer to $100$ days.
In contrast the critically rotating population contains donors with masses only up to ${\sim} 10~M_\odot$, fewer BH companions, a skewed eccentricity distribution peaking around $0.2$, and an orbital period distribution peaking closer to $10$ days.

\begin{figure}[!h]
    \centering
    \includegraphics[width=\columnwidth]{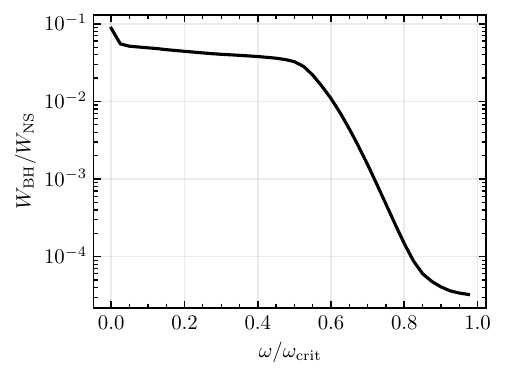}
    \caption{Weighted fraction of \ac{BH} to \ac{NS} Be-XRBs in our population as a function of $\rot{}$, above which we assume the Be phenomenon is efficiently produced.}
    \label{fig:BH_NS_fraction}
\end{figure}

We also vary our choice of threshold rotation ($\rot > 0.5$ in our fiducial model), above which we assume the Be phenomenon is fully efficient.
In \autoref{fig:BH_NS_fraction} we show the impact of this choice on a summary statistic: the weighted number ratio of BH to NS Be-XRBs ($W_\mathrm{BH}/W_\mathrm{NS},$ where $W_\mathrm{CO} = \sum_{i \in \mathrm{bin}} w_{t_\mathrm{det},\mathrm{CO}}^{i}$) as a function of the critical rotation cut, as this ratio is sensitive to the physical channels producing Be-XRBs \citep{Belczynski-Ziolkowski2009}.

Initially, the ratio of BH to NS Be-XRBs is around $\sim 0.3$ with no rotation constraints and monotonically decreases down to about $10^{-4}$ at $\rot \simeq 0.95$.
After an initial plateau, there is a sharp decrease around $\rot \simeq 0.5$, after which $W_\mathrm{BH}/W_\mathrm{NS}$ drops by 2 orders of magnitude.

\section{Discussion}
\label{sec:discussion}

\subsection{Matching with Galactic Be-XRBs}
\label{sec:discussion:comp_to_obs}
Our fiducial BPS modeling and Be-XRB selection criteria show good agreement with the orbital properties and X-ray luminosities of the observed sample of Galactic Be-XRBs.
However, there are some notable discrepancies between our fiducial model and the observed data: We predict more systems with both low eccentricity and low orbital periods than in the observed sample. 
This apparent overprediction of systems at short orbital periods is seen in other BPS studies as well, within the range of commonly explored parameters \cite[e.g. \ac{RLO} \ac{MT} efficiency or \ac{SN} model][]{Shao_Li2014ApJ,Vinciguerra2020}.
Furthermore, tidal truncation of the decretion disk may explain the paucity of short-period Be-XRBs, with an estimated minimum at $P_\mathrm{orb} \simeq 3$ days \citep{Panoglou2018MNRAS}, which is in good agreement with our models as seen in \autoref{fig:period_dist_Fortin}.
Future studies with \posydon{} may explore the contribution between selection effects and binary evolution physics in forming the short-period Be-XRB population.

\subsubsection{On selection effects}\label{sec:observed_sample:selection_effects}
In this work, we only perform qualitative comparisons to the observed Galactic Be-XRB population because we do not account for all observational selection effects, which could induce apparent differences between the observed and predicted system parameters. 
Notably, in \autoref{fig:HMXB_population_p_vs_e_data}, some orbital configurations which have support in our model do not have corresponding observed systems with full orbital solutions. 
Importantly, though, our fiducial model does produce some support for the values of all observed Be-XRB systems, including the nearly circular systems at $P_\mathrm{orb} \simeq 100$ days. 
Future investigations may implement more complete selection effects, though we note that these effects are complex: X-ray activity duty cycles (folding in decretion disk growth/dissipation \citep{Vieira2017MNRAS} and details of accretion efficiency \citep{Brown2018MNRAS}), donor characterization (dependent on luminosity estimates and associations/field crowding; \citealp{Laycock2010ApJ}), and successful eccentricity measurement. 
With a full understanding of the selection effects of our Be-XRB catalog, a comprehensive parameter study comparing observations with our models from \posydon{} may also be performed.

\subsection{Comparison of Be-XRB selection criteria: rotation and donor mass}
\label{sec:discussion:rot_and_mass_cuts}
A key finding of our simulations is that the critical rotation criterion ($\rot{} > 0.5$) is primarily responsible for matching a majority of features in the observed Be-XRB population.
Rotation naturally selects binaries with donor masses in the range $\sim 3-25 ~M_\odot$, which is within the canonical B to late O-type mass range \citep{Harmanec1988BAICz}.
Donor stars with $M_\mathrm{d} \lesssim 2~M_\odot$ are excluded due to magnetic braking efficiently spinning them down, while massive O stars with $M_\mathrm{d} \gtrsim 25-30~M_\odot$ are excluded due to their higher wind mass-loss rates, efficiently removing spin angular momentum from the star.
This is consistent with effects observed in studies focusing on the emergence of the Be phenomenon in single stars \citep{Meynet-Maeder2000,Hastings2021}.

In terms of orbital properties, high rotation selects systems which are preferentially eccentric, as systems which are initially eccentric from the \ac{SN} and circularize from tides necessarily have the stellar spin synchronized ($\tau_\mathrm{synch}<\tau_\mathrm{circ}$, \citealt{Hut1981A&A,Burkart2012MNRAS}), and for the standard Be donors observed in Be-XRBs, their synchronized rotation rates are well sub-critical\footnote{For example, a tidally synchronized star of $15~M_\odot$ with radius $20~R_\odot$ in a $5$ day orbit has a rotation rate of only $\rot{} \simeq 0.1$, well outside the range for observed Be stars \citep{Cranmer2005ApJ,Zorec2016AA,Balona-Ozuyar2021ApJ}.}.
This is consistent with the subset of Be-XRBs with complete orbital solutions, showing a preference for non-circular binaries \citep{Fortin2023AA}.
While the majority of observed Be-XRBs have orbital periods $\gtrsim 10$ days, we see that rotational evolution alone is not sufficient to match this lower limit.
This may be due to enhanced gravitational truncation effects that limit the formation of the decretion disk at short orbital periods \citep{Okazaki2002MNRAS,Panoglou2018MNRAS}.

Finally in the \ac{CO} mass distribution, rapid rotation selects a majority of systems containing NSs relative to BHs, consistent with the observed sample being almost entire \ac{NS} thus far \citep{Reig2011}, and previous theoretical studies explaining the missing BH Be-XRB problem \citep{Belczynski-Ziolkowski2009}.
We find this is primarily driven by the relative rarity of BH progenitor stars and their prior evolution during the HMS-HMS phase.

\subsection{Mass-transfer history}
\label{sec:discussion:MT_history}
Overall, our BPS models are in agreement with the standard formation scenario for Be-XRBs, finding that essentially all systems form through dynamically stable \ac{RLO} \ac{MT} interactions prior to the first SNe \citep[][]{Rappaport_vdHeuvel1982IAUS,Portegies-Zwart1995AA,Pols1991AA}.
More than half of our Be-XRBs go through a phase of \ac{RLO} \ac{MT} initiated at TAMS of the primary followed by \ac{MT} from a stripped helium companion (Case BB), consistent with previous BPS studies \citep[e.g.][]{Shao_Li2014ApJ,Vinciguerra2020}.

Our rotationally-limited accretion formalism strongly influences the outcome of \ac{RLO} \ac{MT}, as stars which accrete only a small fraction of their initial mass are spun up to large fractions of their breakup velocity \citep{Packet1981AA}.
A consequence of this modeling choice manifests in both the implied mass transfer efficiency ($\bar{\beta_\mathrm{rot}}$, \S\ref{sec:results:characteristic_be_XRB_evol}) which is a focus of many studies in binary stellar evolution \citep[e.g.][]{de_Mink_Pols_Hilditch2007AA}, and the ejected circumstellar material from the enhanced wind of the accretor.
This ejected circumstellar material 
can interact with the primary's subsequent \ac{SN}, which has been invoked to explain the light curves of hydrogen-poor superluminous supernovae \citep[][]{Stevance_Eldridge2021MNRAS}, or \acp{SN} with multiple peaks in their light curves \citep[][]{Li2020ApJ}.
Thus, in future studies with \posydon{} we may explore the correlation between post-\ac{MT} binary populations and the occurrence of circumstellar interacting \acp{SN}.

Another predicted population formed through \ac{MT} interactions include stripped helium stars with Be companions \citep[sdO/Be binaries; e.g.][]{Shao_Li2014ApJ,Shao2021ApJ}.
Although challenging to detect \citep{Ramachandran2023AA}, observed sdO/Be binaries have been used to derive \ac{MT} efficiencies, often suggesting highly conservative \ac{MT} \citep{Schootemeijer2018AA,El-Badry2022MNRAS}.
While some of our models undergo nearly conservative \ac{MT}, they represent a negligible fraction of the total population, which instead evolve through highly non-conservative \ac{MT}.
However, it is challenging to make a direct comparison as not all sdO/Be binaries enter an XRB phase (which we consider in this study), as the leftover He core is not always massive enough to form a \ac{CO} and remain bound after the \ac{SN}.
Therefore, a dedicated investigation is needed to compare these populations in the future.

\subsection{Rotational evolution of Be stars}
\label{sec:discussion:rotation_evolution}
If, as a large body of evidence suggests \citep[e.g.][]{El-Badry_Quataert2021MNRAS,El-Badry2022MNRAS,Dodd2023MNRAS}, the majority of Be stars are the result of binary interactions, then we can gain insights into the Be phenomenon by comparing our results to observations of Be stars at large.
Remarkably, our models successfully reproduce multiple features of the observed rotational distribution of Be stars.

We first exclude the slowly rotating population ($\rot{} \simeq 0$ in \autoref{fig:omega_evolution}) from comparison, as we expect moderate rotation is required for stars to exhibit the Be phenomenon.
Then the range of rotation rates for moderate to rapidly rotating stars in our population is naturally set by the paucity of stars with $\rot{} \simeq 0.2-0.4$ (compared to their initially flat distribution in the same range pre-SNe), and fast rotators extending up to to near critical rotation rates $\rot{} \simeq 0.99$.
This span of rotation rates in our population matches the range of observed rotation rates of Be stars from $0.3<v/v_\mathrm{crit} <0.95$ \citep{Cranmer2005ApJ,Zorec2016AA,Balona-Ozuyar2021ApJ}.

Considering the morphology of the rotational distribution, observed Be stars exhibit a single peak at $v/v_\mathrm{crit} \simeq 0.65-0.7$ \citep{Cranmer2005ApJ,Zorec2016AA,Balona-Ozuyar2021ApJ}, which is well matched by the intermediate rotation peak in our models at $\rot{} \simeq 0.6$.
However, our population models seem to overproduce critical rotators with $\rot{} \simeq 0.99$ which is likely due to the use of single-star models during detached evolution which cannot account for effects induced by such rapid rotation (e.g. rotationally enhanced winds).

Further, the bimodal rotational distribution in \autoref{fig:bimodal_rot_hist} is well separated in donor mass, where low mass stars (late-type B) have much higher critical rotation rates than their more massive counterparts (early-type B) which is in agreement with observations of the mean $v/v_\mathrm{crit}$ increasing for later spectral type stars \citep{Yudin2001AA,Cranmer2005ApJ,Balona-Ozuyar2021ApJ}.

We expect the morphology of the rotational distribution to change with our assumed binary stellar evolution physics \citep[e.g.][]{Jagadeesh2023RAA}, metallicity \citep[e.g.][]{Liu2024MNRAS,Misra2023AA}, and adopted star formation history, making measurements of Be star rotational distributions in diverse environments of high interest.
We leave a thorough parameter study predicting said rotational distributions of Be stars in binaries to future work.

\subsection{Donor masses of Be-XRBs}
\label{sec:discussion:donor_masses}
The spectral type distribution of Be stars in star clusters is observed to be bimodal with peaks at early ($\sim$B0-B2) and late ($\sim$B5-B8) spectral types  \citep[e.g.][and references therein]{Mermilliod1982AA,Mathew2008MNRAS,Blesson2008MNRAS,Yu.P.2015AJ,Aidelman2018AA,Jagadeesh2023RAA}.
However, late-type Be stars are not present in the Galactic Be-XRB donor population, which is firmly peaked at early spectral types (B0-B2) corresponding to about $M_\mathrm{d} \simeq 15~M_\odot$ \citep{Fortin2023AA}.
This lack of low mass donors in Be-XRBs is often attributed to the stability and/or efficiency of \ac{RLO} \ac{MT}, that preferentially disfavors their formation \citep[][]{Shao_Li2014ApJ,Vinciguerra2020,Igoshev2021MNRAS}.
Reproducing the mass distribution of Be-XRBs has long been recognized as a challenge for theoretical models \citep{Portegies-Zwart1995AA}, and is often considered an open question in BPS studies.

With our implementation of rotationally-limited accretion in \posydon{}, we have shown in \autoref{fig:implicit_beta_distr} our population exhibits highly non-conservative \ac{MT}, with a \ac{RLO} \ac{MT} efficiency distribution peaking around $\bar{\beta}_\mathrm{rot} \simeq 5\%$.
When comparing to other BPS studies which explore an equivalently low value of the \ac{MT} efficiency, we produce similar results, finding a donor mass distribution which peaks around $M_\mathrm{d} \simeq 3-5~M_\odot$  \citep[e.g.][]{Shao_Li2014ApJ,Vinciguerra2020,Igoshev2021MNRAS}, in tension with the observed Be-XRB donor mass distribution.
One way the aforementioned BPS studies successfully match the observed donor mass distribution is by increasing the \ac{MT} efficiency of all binary interactions -- a modeling choice we cannot trivially adopt in our \posydon{} binary \mesa{} models.
Thus, our fiducial BPS model is unable to reproduce the observed Be-XRB donor mass distribution.
However, we maintain the improved physical accuracy and interpretation of our \ac{MT} efficiency distribution.

If \ac{MT} efficiency is well described by our rotationally-limited accretion models, our study suggests another source of preferential selection disfavoring lower mass donors in Be-XRBs.
This may be due to selection effects not yet considered from the Be phenomenon including its dependence on spectral type or evolutionary state \citep[e.g. Be vs.\ Bn stars;][]{Mathew2008MNRAS,Cochetti2020AA}, and measured disk properties (density, duty cycle, and/or emission physics; \citealt{Vieira2017MNRAS}).

If, on the other hand, \ac{MT} efficiency is more conservative than our detailed models suggest, our present study demonstrates the need for new treatment beyond our rotationally-limited accretion formalism, marking a significant step toward a physical interpretation of the \ac{MT} efficiency in BPS studies.
For example, in the future, we may incorporate additional angular momentum loss mechanisms into \posydon{} models, such as anisotropic winds due to gravity darkening \citep[][]{Georgy2011AA}, angular momentum transport from the decretion disk \citep[][]{Krticka2011AA}, and/or the winds launched from the disk itself \citep[][]{Ressler2021MNRAS}.

\subsection{On the dearth of BH Be-XRBs}
\label{sec:discussion:BH_NS_BeXRBs}
We find a significantly higher frequency of NS Be-XRBs, suggesting BH Be-XRBs are much less likely to be in the intrinsic Galactic Be-XRB population, primarily due to MT interactions and the rarity of BH progenitor stars.
This is consistent with observations finding nearly all Be-XRBs with characterized \ac{CO} companions harbor a \ac{NS} \citep{Reig2011}, with one tentative BH Be-XRB claimed \citep[MWC-656;][however \citealp{Janssens2023AA} favor a \ac{NS} or less massive companion]{Casares2014Nature}.

While it has been shown that the prevalence of \ac{NS} hosting Be-XRBs is sensitive to the adopted binary stellar evolution physics including \ac{RLO} \ac{MT} \citep[e.g.][]{Belczynski-Ziolkowski2009,Shao_Li2014ApJ}, we have demonstrated in \autoref{fig:BH_NS_fraction}, the parameterization of the Be phenomenon itself can have a comparably large impact on the ratio of \ac{BH} to \ac{NS} Be-XRBs.
This suggests that inferences drawn from Be-XRB BPS studies may be obfuscated by the uncertainties underlying the Be phenomenon, highlighting the importance of theoretically studies reproducing the Be phenomenon self-consistently \citep[e.g.][]{Ressler2021MNRAS}, and predicting rotational distributions of Be stars to compare with observations.

Moreover, the number ratio of \ac{BH} to \ac{NS} Be-XRBs (\autoref{fig:BH_NS_fraction}) should be considered an upper limit as accretion luminosity selection effects will down-weight \acp{BH} compared to \acp{NS} in similar accretion conditions \citep{Brown2018MNRAS}.
Since the number ratio of \ac{BH} to \ac{NS} Be-XRBs is a strong function of the threshold rotation chosen for the Be phenomenon, the discovery of BH Be-XRBs may provide valuable constraints on the Be phenomenon itself and binary stellar evolution models.

\subsection{Caveats}
\label{sec:discussion:caveats}
To compare with the Galactic sample, we adopt models at solar metallicity and use a constant star formation history, although it has been shown the Milky Way has a more complex metallicity distribution and star formation history \citep[e.g.,][]{Snaith2015AA}.
Since this is the first study to explore the self-consistent rotational evolution of Be-XRBs along their HMS-HMS evolution, we use these approximations as these are consistent with previous BPS studies \citep[e.g.][]{breivik2017} and serve as a benchmark for comparison for future studies.
Moreover, the complex selection effects in HMXB analyses strongly influence our predicted distribution of Be-XRBs, which may be comparable deviations introduced by changing our fiducial assumptions about the Milky Way.

In our fiducial Be-XRB models, we use a conservative\footnote{Compared to other BPS studies which use a wide range of reported values that are generally at higher cutoffs of $\rot{} > 0.7-0.8$ or greater \citep[e.g.][]{Shao_Li2014ApJ,Misra2023AA,Liu2024MNRAS}, a result of the confusion within the literature about the true rotation rates of Be stars.} critical rotation cutoff of $\rot>0.5$, where we assume the Be phenomenon is fully efficient while the star is above the threshold.
However, it is well known the Be phenomenon is inherently transient, with a duty cycle linked to the hereto unknown disk launching mechanism(s) \citep[e.g. Be vs.\ Bn stars;][]{Cochetti2020AA}, and the structural evolution directly observed in Be star disks \citep{Vieira2017MNRAS}.
Therefore a complete treatment of Be-XRB observables must include more detailed treatment of disk formation scenarios.
However, as we have shown, even with these simplifying assumptions we are able to match a substantial number of observable properties of the Galactic Be-XRB population.

When considering the bimodal rotation distribution explored in \autoref{fig:individual_param_hist} and \autoref{fig:bimodal_rot_hist}, we note that modeling the detached phase is performed using pre-computed single-star models which cannot self-consistently evolve the stellar structure in response to rapid rotation.
Since rotational effects may significantly alter the evolution of stars rotating close to their critical limit \citep[][]{Georgy2011AA}, we expect the peak present in our models at $\rot \approx 0.99$ to be shifted and or smoothed out to lower values of $\rot$ which would be further in line with observational studies emphasizing the paucity of near critically rotating Be stars \citep[][]{Zorec2016AA,Balona-Ozuyar2021ApJ}

We do not vary uncertain parameters (as is standard in BPS studies) such as the common-envelope efficiency \citep[][]{Belczynski-Ziolkowski2009}, or SNe models and kick prescriptions \citep[][]{Igoshev2021MNRAS}, as we expect these uncertainties to be comparable to those introduced in modeling the Be phenomenon itself and other complex selection effects present for HMXBs.
Furthermore, we expect the physical mechanisms forming the rotational distributions highlighted in this study to be robust against such choices.

While \posydon{} v1 is unable to model systems which go through a successful CE with two H-rich stars followed by a second phase of MT, \citealt[][]{Fragos2023ApJS} have shown the fraction of systems evolving through this channel is negligible ${\sim}0.3\%$, and thus, do not impact the present study.
\section{Conclusions}
\label{sec:conclusions}
We have modeled Be X-ray binaries (Be-XRBs) using the \posydon{} binary population synthesis (BPS) code \citep{Fragos2023ApJS} performing a detailed study of the population-level spin evolution of stars undergoing \acf{RLO} \acf{MT}, through their detached XRB phase.
We use binary \mesa{} models at solar metallicity combined with a constant star formation history, to compare with properties of Galactic Be-XRBs from a select subset of the catalogue compiled by \cite{Fortin2023AA}.
We employ the unique modeling capabilities of \posydon{} to investigate the role rotation has as a selection criterion for identifying Be-XRB-like systems from our synthetic models.
We find that using rapid rotation ($\rot{} > 0.5$) as a selection criterion naturally reproduces many features of the observed Galactic Be-XRB population self-consistently.

Our fiducial BPS models in \posydon{} are able to reproduce the broad features of the orbital properties of the observed Galactic Be-XRB population. 
We reconfirm the standard evolution pathway forming Be-XRBs (stable \ac{MT}) and strengthen the binary formation scenario for Be stars in general, where we have demonstrated the importance of single star effects (structural changes towards TAMS and mass dependent winds) and tidal evolution in forming the rotational distribution during detached evolution.

We find the rotational distribution of our model Be-XRBs contains many features in agreement with the literature reported values of Be stars spinning in the range $0.3 \lesssim v/v_\mathrm{crit} \lesssim 0.95$ with a peak around $v/v_\mathrm{crit} \simeq 0.65$ \citep{Zorec2016AA,Balona-Ozuyar2021ApJ}. 
While non-radial stellar pulsations have been shown to successfully form sustained decretion disks for near critical rotators ($\rot{} \gtrsim 0.95$), the impulsive magnetic rotator model may be a dominant formation mechanism for producing the Be phenomenon at this secondary peak at more modest rotation rates $\rot{} \simeq 0.6$ and below \citep{Balona-Ozuyar2021ApJ}.

With the detailed binary simulations from \posydon{}, we calculate a rotationally-limited mass-transfer efficiency, $\bar{\beta}_\mathrm{rot}$, and consider the population-level distribution of \ac{MT} efficiencies.
Our models suggest the majority of systems evolve through highly non-conservative MT with an equivalent $\bar{\beta}_\mathrm{rot} \approx 0.05$.
This confirms the picture that a significant fraction of Be-XRBs evolve through a sdO/Be phase before the first \ac{SN}, and are therefore prime candidates for exhibiting circumstellar interacting SNe.

When considering different features in the rotational distribution of our Be-XRBs, our results suggest that \ac{BH} Be-XRBs are more likely to contain higher mass donors with moderate rotation rates ($\rot \simeq 0.6$), and be in relatively wider orbits ($P_\mathrm{orb} \simeq 100-1000 ~\mathrm{d}$) with a uniform eccentricity distribution.
Furthermore, we find that the parameterization of the Be phenomenon itself (the critical rotation threshold) can impact the resulting Be-XRB population at a level comparable to changing the adopted binary stellar evolution physics (e.g. \ac{SN} models or the efficiency and stability of \ac{RLO} \ac{MT}).

In future studies we will explore the  metallicity dependence of the rotational distribution of donors in Be X-ray binaries and its dependence on the star formation history in environments such as the Small Magellanic Cloud.
With this study, we are now poised to explore more physically motivated mass-transfer prescriptions for binary evolution, where Be-XRBs may be a key population for constraining the physics of \ac{RLO} \ac{MT} in interacting binaries.
With the ever increasing complexity of Be-XRB models, we may soon uncover the mechanisms responsible for the Be phenomenon, allowing further investigation into open questions in BPS such as CO formation and SNe physics.

\section*{Acknowledgments}
The POSYDON project is supported primarily by two sources: the Swiss National Science Foundation (PI Fragos, project numbers PP00P2\_211006 and CRSII5\_213497) and the Gordon and Betty Moore Foundation (PI Kalogera, grant award GBMF8477).
K.A.R.\ is supported by the Gordon and Betty Moore Foundation (PI Kalogera, grant award GBMF8477) and the Riedel Family Fellowship. 
K.A.R.\ also thanks the LSSTC Data Science Fellowship Program, which is funded by LSSTC, NSF Cybertraining Grant No.\ 1829740, the Brinson Foundation, and the Moore Foundation; their participation in the program has benefited this work.
J.J.A.~acknowledges support for Program number (JWST-AR-04369.001-A) provided through a grant from the STScI under NASA contract NAS5-03127.
Z.D.\ acknowledges support from the CIERA Board of Visitors Research Professorship. 
S.S.B., T.F., and Z.X. were supported by the project number PP00P2\_211006. 
S.S.B. was also supported by the project number CRSII5\_213497. 
Z.X. acknowledges support from the Chinese Scholarship Council (CSC).
K.K. acknowledges support from the Spanish State Research Agency, through the María de Maeztu Program for Centers and Units of Excellence in R\&D, No. CEX2020-001058-M. 
E.Z. acknowledges funding support from 
the Hellenic Foundation for Research and Innovation (H.F.R.I.) under the ``3rd Call for H.F.R.I. Research Projects to support Post-Doctoral Researchers" (Project No: 7933). 
This research was supported in part through the computational resources and staff contributions provided for the Quest high performance computing facility at Northwestern University which is jointly supported by the Office of the Provost, the Office for Research, and Northwestern University Information Technology.
K.A.R.\ thanks Tomer Shanar, Yoram Lithwick, and Giles Novak for insightful discussions on interpreting observations and accretion disk physics. 
We thank the referee for providing valuable feedback improving the quality of this work.

\software{\texttt{NumPy} \citep{numpy}, {\tt SciPy} \citep{scipy:2020NatMe..17..261V}, {\tt matplotlib} \citep{matplotlib}, {\tt pandas} \citep{pandas-mckinney-proc-scipy-2010}, \posydon{} \citep{Fragos2023ApJS}}

\bibliographystyle{aasjournal}
\bibliography{references.bib}

\appendix
\section{Supplementary Material}
\label{sec:appendix}
\restartappendixnumbering

\begin{figure*}[hbp]
    \centering
    \includegraphics[width=\columnwidth]{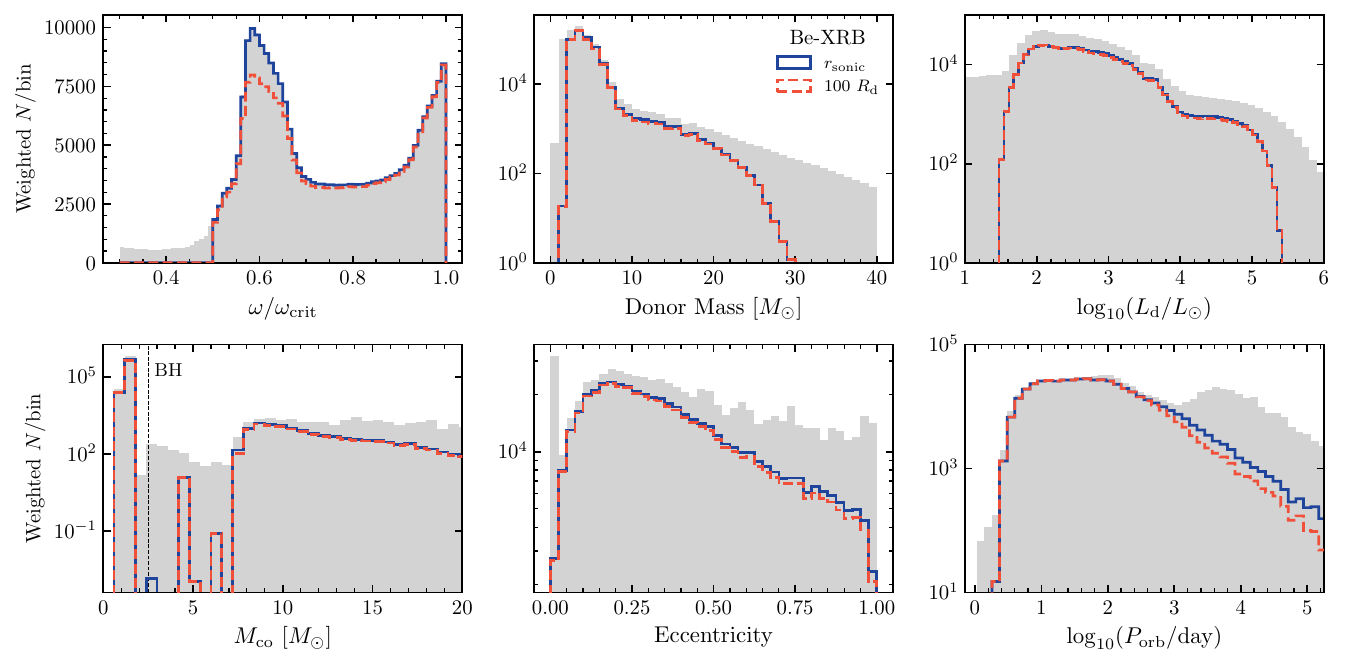}
    \caption{One-dimensional parameter histograms  comparing the baseline (gray) and Be-XRB population with two different disk models ( \S\ref{sec:methods:disk_interaction}), weighted by their detached lifetime ($w_{t_\mathrm{det}}$). The interaction criterion compares the decretion disk radius to the periapse separation of the binary to determine systems which will produce X-ray emission. The strong overlap in all parameters demonstrates the resulting Be-XRB population is robust to the parameterization of the disk.}
    \label{fig:comparing_disk_models}
\end{figure*}

\autoref{fig:comparing_disk_models} shows the baseline and Be-XRB subpopulations with two different decretion disk models (Section \ref{sec:methods:disk_interaction}) which demonstrate our results are robust to the adopted disk model.
We expect other factors such as the decretion disk duty cycle and base gas density to be leading uncertainties in future studies modeling the observable Be-XRB population, as these parameters are directly coupled to the predicted X-ray luminosity/lifetime and are highly uncertain.
\autoref{table:Fortin_BeXRB_params} shows all observational data of Galactic Be-XRBs used in this study, taken from the catalog compiled by \citealt{Fortin2023AA}.
\autoref{table:mass-transfer-cases} shows the formation efficiency of our baseline and Be-XRB population compared to our initial population of $2\times10^6$ binares, as a function of their MT history. The last columns indicate the fraction of Be-XRBs (weighted and unweighted) which evolved through a given MT history.

\begin{deluxetable*}{c c c c c}
    \tablecaption{Be-XRBs from the catalogue compiled by \cite[][with refernces therein]{Fortin2023AA}, keeping only the systems with Be designations, and measured periods and/or eccentricities.
    \label{table:Fortin_BeXRB_params}
    }
    \setlength{\tabcolsep}{20pt}
    \tablehead{
    \colhead{ID} & Spectral Type & $P_\mathrm{orb}$ [day] & Eccentricity  & Used for $e-P_\mathrm{orb}$
    }
    \startdata
    AX J163904-4642&BIV-V&4.23785&-&False \\
    EXMS B1210-645&B2V&6.7&-&False \\
    4U 2206+543&O9.5Vep&9.558&0.3&True \\
    SAX J0635.2+0533&B2V-B1IIIe&11.2&0.29&True \\
    SAX J2103.5+4545&B0Ve&12.66536&0.4055&True \\
    Cep X-4&B1-B2Ve&20.85&-&False \\
    4U 1901+03&B8-9 IV&22.5827&0.0363&True \\
    1A 1118-615&O9.5III-Ve&24.0&-&False \\
    4U 0115+634&B0.2Ve&24.3174&0.339&True \\
    Swift J0243.6+6124&O9.5V&28.3&0.092&True \\
    H 1553-542&B1-2 V&30.6&-&False \\
    3A 0726-260&O5Ve&34.548&-&False \\
    V 0332+53&O8.5Ve&36.5&0.417&True \\
    XTE J1859+083&B0-2Ve&37.97&0.127&True \\
    KS 1947+300&B0Ve&40.415&0.034&True \\
    H 1417-624&B1e&42.12&0.446&False \\
    AX J1700.2-4220&B0.5IVe&44.03&-&False \\
    RX J2030.5+4751&B0.5V-IIIe&46.02&0.41&False \\
    EXO 2030+375&B0Ve&46.0214&0.419&True \\
    IGR J14488-5942&O-BVe&49.63&-&False \\
    IGR J11435-6109&B0.5Ve&52.46&-&False \\
    AX J1820.5-1434&B0-2 IV-Ve&54.0&-&False \\
    GRO J2058+42&O9.5-B0IV-Ve&55.0&-&False \\
    MWC 656&B1.5-B2IIIe&60.37&0.1&True \\
    4U 1036-56&B0III-Ve&60.9&-&False \\
    RX J0812.4-3114&B0.2IVe&80.39&-&False \\
    HD 119682&B0Ve&90.0&-&False \\
    3A 0656-072&O9.7Ve&101.2&-&False \\
    GS 0834-430&B0-2 III-Ve&105.8&0.14&True \\
    1A 0535+262&O9.5III-Ve&110.3&0.47&True \\
    IGR J19294+1816&B1Ve&117.2&-&False \\
    IGR J11305-6256&B0IIIe&120.83&-&False \\
    GX 304-1&B2Vne&132.5&-&False \\
    SWIFT J1626.6-5156&B0Ve&132.89&0.08&True \\
    RX J0440.9+4431&B0.2Ve&150.0&-&False \\
    SWIFT J1816.7-1613&B0-2e&151.1&-&False \\
    IGR J01363+6610&B1Ve&159.0&-&False \\
    XTE J1946+274&B0-1IV-Ve&172.7&0.246&True \\
    AX J1749.1-2733&B1-3V&185.5&-&False \\
    2E 1145.5-6155&B1III-Ve&187.5&0.5&False \\
    1H 1249-637&B0.5IVpe&226.0&-&False \\
    GRO J1008-57&B0e&247.8&0.68&True \\
    X Per&B1Ve&250.3&0.111&True \\
    SAX J2239.3+6116&B0Ve&262.0&-&False \\
    RX J0146.9+6121&B1IIIe&330.0&-&False \\
    PSR B1259-63&O9.5Ve ($\gamma$\,Be)&1236.724526&0.8698797&True \\
    \enddata
\end{deluxetable*}

\begin{deluxetable*}{c c c c}[hbp]
    \tablecaption{Formation efficiency of our CO+star baseline population and Be-XRBs organized by mass-transfer evolution during their HMS-HMS phase. The first column indicates the fraction of systems that enter our baseline population out of the initial samples of $2\times10^6$ binaries, and the fraction of those which eventually become Be-XRBs. The last column indicates the fraction of Be-XRBs that went through a given MT channel. \label{table:mass-transfer-cases}}
    \setlength{\tabcolsep}{20pt}
    \tablehead{
        \colhead{MT case} & Baseline $\rightarrow$ Be-XRB (N) & Be-XRBs (N) & Be-XRBs ($w_{t_\mathrm{det}}$)
    }
    \startdata
        B           & $34.0\% ~ (56.3\%)$ & $33.1\%$ &   $19.7\%$ \\
        B/C/BB      & $20.5\% ~ (86.9\%)$ & $30.8\%$ &   $47.1\%$ \\
        no RLO      & $14.7\% ~ (2.3\%)$ & $0.6\%$  &  $<0.1\%$ \\
        A/B  	    & $11.9\% ~ (44.3\%)$ & $9.1\%$  &    $1.4\%$ \\
        B/C  	    & $7.4\%  ~ (91.5\%)$ & $11.7\%$  &  $14.5\%$ \\
        B/BB 	    & $6.4\%  ~ (94.7\%)$ & $10.5\%$  &  $15.6\%$ \\
        A/B/BB 	    & $1.9\%  ~ (93.5\%)$ & $3.1\%$   &   $1.4\%$ \\
        A    	    & $1.7\%  ~ (14.2\%)$ & $0.4\%$   &   $0.1\%$ \\
        C    	    & $1.5\%  ~ (24.7\%)$ & $0.6\%$   &  $0.2\%$ \\
    \enddata
\end{deluxetable*}

\end{document}